\documentclass[smus]{snow2e}
\usepackage{graphics}
\def\gapprox{\mathrel{\raise .32ex\hbox {$>$}\mkern-14mu \lower0.62ex\hbox{$\sim$}}}
\def\lapprox{\mathrel{\raise .32ex\hbox {$<$}\mkern-14mu \lower0.62ex\hbox{$\sim$}}} 
\def\Emiss{\rlap{E}{/}}
\def\Dzero{D$\emptyset$}
\def\GeV{{\rm GeV}}
\def\lsim{\, \lower0.5ex\hbox{$\stackrel{<}{\sim}$}\, }
\def\gsim{\, \lower0.5ex\hbox{$\stackrel{>}{\sim}$}\, }

\input epsf
\begin{document}
\hyphenation{counter-terms}
\hyphenation{VLHC}
\hyphenation{LHC}
\hyphenation{Beckons}
\begin{titlepage}
\headsep 2cm
\baselineskip 0.5cm
\evensidemargin 0.7cm
\begin{flushright}
{\large FERMILAB-CONF-97/ 318-T}
\end{flushright}
\vspace{.5cm}
\begin{center}
{\LARGE \bf Summary of the Very Large Hadron Collider Physics and Detector Workshop}\\
\vspace{.5cm}
{\Large \it Physics at the high energy frontier beyond the LHC}\\ 
March 13-15, 1997\\
Fermi National Accelerator Laboratory, Batavia, Illinois\\
\vspace{.5cm} 
\end{center}
{
G.~Anderson (Fermilab),
U.~Baur (SUNY at Buffalo), 
M.~Berger (Indiana University), 
F.~Borcherding (Fermilab), 
A.~Brandt (Fermilab), 
D.~Denisov (Fermilab, Co-Chair and Co-editor), 
S.~Eno (University of Maryland), 
T.~Han (University of California--Davis),
S.~Keller (Fermilab, Co-Chair and Co-editor), 
D.~Khazins (Duke University), 
T.~LeCompte (Argonne National Laboratory), 
J.~Lykken (Fermilab), 
F.~Olness (Southern Methodist University),
F.~Paige (Brookhaven National Laboratory), 
R.~Scalise (Southern Methodist University),
E.~H.~Simmons (Boston University), 
G.~Snow (University of Nebraska--Lincoln), 
C.~Taylor (Case Western Reserve University), 
J.~Womersley (Fermilab).}\\

\begin{center}
\vspace{.5cm}
{\Large \bf Abstract}\\
\end{center}
\vspace{.5cm}
{\large 
One of the options for an accelerator beyond the LHC is a hadron 
collider with higher energy.
Work is going on to explore accelerator technologies that would 
make such a machine feasible.  This workshop concentrated on the physics and 
detector issues associated with a hadron collider with an energy in the center 
of mass of the order of 100 to 200 TeV. 
}

\end{titlepage}
\pagestyle{plain}
{\large \section{Introduction}\vfill}
The Very Large Hadron Collider Physics and Detector workshop took
place at Fermilab in March 1997.  In this paper we summarize the
activities of the working groups during the workshop.

This workshop was motivated by the
accelerator work~\cite{PIPE}~\footnote{The references for the
introduction and conclusions are at the end of the paper, the
references for the working group summaries are at the end of the
respective sections.}  that has been started on new technologies for a
post-LHC Very Large Hadron Collider (VLHC).  Obviously, physics and
detector issues, along with accelerator technology and budget
constraint, must guide us to select appropriate and realistic energy
and luminosity for such a machine.

As is well known, the last largely unexplored sector of the \linebreak Standard
Model (SM), the Higgs sector, will be investigated over the next
decade or so by the Tevatron, HERA, LEP, and LHC.  Any post-LHC
machine will be built to explore physics beyond the SM.  At this point
in time, we do not have any experimental evidence for the physics
beyond the SM, and it is therefore difficult to make the case for any
specific accelerator beyond the LHC.  Therefore, our goal is to make
the case for accelerator and detector developments that would allow us
to build a hadron collider for a lower cost than with current technologies.

Some preliminary work was done during the Snowmass 96~\cite{SNOW96}
workshop, where the EHLQ~\cite{EHLQ} paper was used as a guide.
Contrary to what is sometimes assumed, it is not necessary to increase
the luminosity proportionally to the square of the energy.  In fact
for the production of heavy objects, each time the accelerator energy
is increased by a factor of 2, the cross section increases by more
than a factor of 10.  For this to be true, the heavy object has to be
detectable at the lower energy accelerator.  The increase in cross
section is due to the simple fact that the average Bjorken-x probed is
decreased when the accelerator energy is increased and that the parton
distribution function are larger at smaller x.

For this workshop it was therefore decided to concentrate on a center of
mass energy ($E_{cm}$) between 100 TeV and 200 TeV and a luminosity
(${\cal L}$) between $10^{34}cm^{-2} s^{-1}$ and $10^{35}cm^{-2}
s^{-1}$.  These ranges are testing the limits of the detector and
accelerator capabilities and allow one to investigate the tradeoff between
$E_{cm}$ and ${\cal L}$.  The increase in $E_{cm}$ and ${\cal L}$ from
the LHC to the VLHC are about the same as the increase between the Tevatron and
the LHC.  We will see that with these parameters the scales of physics
beyond the SM that can be probed are about an order of magnitude larger than
the scales probed at the LHC.

One of the conclusions reached during the Snowmass 96 workshop was
that it would be interesting to concentrate on scenarios of physics
beyond the SM that have a chance to reveal themselves before the VLHC
and study their implications for a VLHC.  This was done in several
studies during this workshop.

Due to the discovery nature of the VLHC, it is clear that we need to
consider multipurpose detectors.  No major problems with detector
design were uncovered during the Snowmass 96 study.

About 150 participants attended the workshop which consisted of one
day of plenary talks, one day of working group meetings and one half
day of working group summaries by the conveners.  Before going to the
summary of the different working groups we provide here a brief
synopsis of the plenary talks.

P. Sphicas (MIT) started with a broad sweep from the first hadron
collider, the ISR, to the VLHC.  He summarized the impresssive
achievements of hadron colliders over the last 25 years and the
current outstanding questions.  G. Dugan (Cornell University)
summarized the activity of the accelerator group during the Snowmass
96 workshop.  Both the low and high field options were considered,
with a center of mass energy of 100 TeV, a peak luminosity of
$10^{34}cm^{-2} s^{-1}$ and 16ns bunch spacing.  It is currently
believed that there is an overall luminosity upgrade potential for the
low and high field options of a factor of 5 and 10, respectively.
P. Bloch (CERN) reviewed the main components of the LHC detectors.
C. Hill (Fermilab) summarized the basic principles of the SM and
possible scenarios of physics beyond the SM (e.g., SUSY, Technicolor,
Topcolor.).  F. Wilczek (IAS, Princeton) gave his view of the future
of particle physics, describing the success and deficiencies of the SM
and the wonders of unification and SUSY.  P. Grannis (SUNY Stony
Brook) closed the workshop with a talk entitled:" The Future Beckons".
He analysed lessons of the past and drew morals for future decisions.
He presented both depressing and optimistics scenarios, and warned us
against the danger of hasting our decision to support a specific
machine before Nature gives us some clues of what the appropriate
machine is.~\footnote{The collection of transparencies of the plenary
talks and working group summary talks is available, please send your
request for a copy to vlhc@fnth32.fnal.gov.}

We now turn to the working group summaries.

\clearpage
\section{New Strong Dynamics Working Group}

\begin{center}
{Elizabeth~Simmons\\
{\textit{Boston University}}\\
John~Womersley\\ 
{\textit{Fermi National Accelerator
Laboratory}}}

\end{center}

The New Strong Dynamics working group considered what a VLHC could
reveal about new strong interactions, such as might be involved in
electroweak symmetry breaking (EWSB).  We tried to identify new
physics that would be uniquely visible at a VLHC (as opposed to the
LHC, NLC, or a muon collider).  We also considered the appropriate
center of mass energy and luminosity for a hadron collider intended to
explore this physics and whether the traditional `rules of thumb'
about energy-luminosity trade-offs hold.

The working group met for a total of four hours during the VLHC workshop.
Group discussions were initiated (and ultimately
summarized) by the following presentations: 
\begin{itemize}
\item Introduction and Overview (J. Womersley)
\item Non-Standard Higgs (V. Koulovassilopoulos)
\item Multiple W Production (W. Kilgore)
\item Strong WW Scattering (K. Cheung)
\item Deca-TeV Unified Compositeness (Y. Pirogov)
\item Summary (E. Simmons)
\end{itemize}
\medskip
\noindent{Many other physicists including S. Chivukula, P. Grannis, 
C.~Hill, T.~LeCompte and F. Paige also made valuable contributions.}

One thread of our discussion centered around the feasibility of using
the VLHC to study a `non-standard Higgs': a scalar boson with a mass
of 400 to 800 GeV with non-standard couplings to weak gauge bosons and
fermions\cite{kk}.  Looking in the decay channel $H \to Z Z \to \ell^+
\ell^- \ell^+ \ell^-$, it appears that this scalar can be discovered
as easily as a standard Higgs.  A careful measurement might then make
it possible to distinguish whether the width of the discovered object
differed from that of the standard model prediction by more than a few
percent.  Both the discovery and identification capabilities of a VLHC
would be superior to those of the LHC; a muon collider of the right
energy might also do a reasonable job.  The relatively low mass of
this scalar makes it easier to study at a lower-energy (60-100 TeV)
VLHC than at a higher-energy (200 TeV) machine.

In particular, Koulovassilopoulos {\it et al.} studied the possibility
that the $WWH$ coupling is rescaled relative to the standard model
value by a factor $\xi$.  The decay width of the heavy Higgs is
proportional to $\xi^2$, and a collider's ability to detect the
non-standard nature of the Higgs can be described in terms of its
sensitivity to deviations of $\xi$ from 1.0.  Their results for the
LHC and several possible VLHC accelerators are listed in Table~I.

\begin{table}
\label{tab:sens}
\begin{center}
\begin{tabular}{|c|c|c|c|}
\hline\hline &\multicolumn{3}{c|}{Sensitivity to $\xi$}\\ \cline{2-4}
$\sqrt{s}$, ${\cal{L}}$(cm$^{-2}s^{-1}$)
&\multicolumn{3}{c|}{$m_H$~(GeV)}\\ \cline{2-4} &400&600&800\\ \hline
\hline 14~TeV, 10$^{33}$ &60\%$^*$&---&---\\ 14~TeV, 10$^{34}$
&20\%$^*$&40\%$^*$&---\\ 50~TeV, 10$^{34}$ &7\%&12\%&20\%\\ 50~TeV,
10$^{35}$ &3\%&4\%&7\%\\ 100~TeV, 10$^{34}$ &6\%&8\%&12\%\\ 100~TeV,
10$^{35}$ &$2-3$\%&3\%&5\%\\ 200~TeV, 10$^{34}$ &---&25\%&30\%\\
200~TeV, 10$^{35}$ &---&8\%&12\%\\ \hline\hline
\end{tabular}
\end{center}
\caption{Sensitivity to the parameter $\xi$ at the LHC and VLHC for
various value of the luminosity and CM energy.  The starred entries
indicate that the value given applies only for $\xi > 1$, whereas for
$\xi < 1$ the sensitivity is substantially worse.}
\end{table}

Another topic was production and detection of multiple (longitudinal)
weak gauge bosons at high energies.  The idea is that just as pion
scattering above the $\rho$ resonance is dominated by multiple pion
production, so might $W_L W_L$ scattering (in a strongly-interacting
regime) result at high energies in multi-$W_L$ final states.
Kilgore\cite{kilgore} estimates the total cross section $\sigma_{WW}$
to be $\sim 100$~fb at the ``$\rho$'' peak and $\sim 20$~fb
asymptotically at higher $\sqrt{\hat s}$.  If the acceptance is of
order 50\%, then the observable cross section of $\sim 10$~fb would
imply that it is not reasonable to require more than $\sim 1$ of the
$W$'s to decay leptonically.  If one of the W's is required to decay
leptonically and each of the others becomes a single `fat' hadronic
jet, the dominant background will arise from standard production of
one W plus multiple gluon jets.  Unlike multiple $W$ production the
cross section for this process would fall rapidly with increasing jet
multiplicity and the multiple-$W$ signal might become apparent above
$n_{\rm jets} \sim 5-8$.  The current rough estimate is that with an
integrated luminosity of 100~fb$^{-1}$, about 100 signal events (and
almost no background) might remain after all cuts and branching ratios
are included.  Other suggestions for reducing background and improving
signal included allowing more than one $W$ to decay leptonically or
identifying tau leptons resulting from $W \to \tau \nu_\tau$ decays.

\begin{table}[t]
\label{tab:signal}
\begin{center}
\begin{tabular}{|c|c|c|c|}
\hline\hline
Mode&$\sigma_B$&\multicolumn{2}{c|}{$\sigma_S$($S/\sqrt{B}$)}\\
\cline{3-4}
&&1 TeV Higgs&Rescaled $\pi\pi$\\
\hline
\multicolumn{4}{|c|}{$\sqrt{s}=14$~TeV, 100~fb$^{-1}$}\\
\hline
$W^\pm W^\pm$&0.037&0.065(3.4)&0.2(11)\\
$ZZ         $&0.006&0.042(5.1)&0.0085(1.0)\\
$W^+ W^-    $&0.13&0.18(5.0)&0.05(1.3)\\
$W^\pm Z    $&0.05&0.016(0.7)&0.04(1.8)\\
\hline
\multicolumn{4}{|c|}{$\sqrt{s}=60$~TeV, 100~fb$^{-1}$}\\
\hline
$W^\pm W^\pm$&0.83&1.4(15)&4.9(54)\\
$ZZ         $&0.2&1.3(29)&0.3(6.9)\\
$W^+ W^-    $&5.2&7.3(32)&3.5(15)\\
$W^\pm Z    $&1.1&0.38(3.6)&1.5(14)\\
\hline
\multicolumn{4}{|c|}{$\sqrt{s}=100$~TeV, 100~fb$^{-1}$}\\
\hline
$W^\pm W^\pm$&1.9&2.9(21)&10(72)\\
$ZZ         $&0.5&2.8(40)&0.68(9.6)\\
$W^+ W^-    $&12&16(45)&9.4(27)\\
$W^\pm Z    $&3.2&0.7(3.9)&4.0(22)\\
\hline
\multicolumn{4}{|c|}{$\sqrt{s}=200$~TeV, 100~fb$^{-1}$}\\
\hline
$W^\pm W^\pm$&4.9&6.8(31)&2.4(110)\\
$ZZ         $&1.3&6.2(52)&1.8(15)\\
$W^+ W^-    $&40&40(62)&28(43)\\
$W^\pm Z    $&10&3.1(6.6)&15(47)\\
\hline
\hline
\end{tabular}
\end{center}
\caption{Signal and background cross sections $\sigma_S$ and 
$\sigma_B$ (in femtobarns), and signal significance $S/\sqrt{B}$, for two
models of vector-boson pair production in various final
states, for the LHC and for various VLHC options.}
\end{table}

A third focus was on how well the VLHC compares with the LHC in
studying strong $V_L V_L$ scattering in the gold-plated modes where
the vector bosons decay leptonically and the silver-plated modes where
two $Z$ bosons are produced and one decays to neutrinos\cite{kingman}.  
Since (at
tree level) the
only hadronic activity in the detector in the signal events would
result from the spectator quarks that radiated the $W_L's$, a forward
jet tag and central jet veto can reduce background.  It appears that
the VLHC would do much better than the LHC at detecting the simple
excess of $V_L V_L$ final states that would indicate the presence of
a strongly-coupled electroweak symmetry-breaking sector. 
Furthermore, the VLHC would more clearly determine which specific
final states ($W^\pm W^\pm, Z Z, W^+ W^-, W Z$) showed the largest
excesses -- information that would help distinguish among competing
models of the strong electroweak interactions.  It is interesting to note 
that, based on these studies, cutting on measured missing $E_T$ 
may not be required for $W$ identification.  This is potentially 
important for detector design. 

For instance, Cheung {\it et al.} examined the signal of a strong 
EWSB sector at the LHC and VLHC in several longitudinal di-boson
final states.  They employed the general type of cuts described above
and evaluated the significance  $S/\sqrt{B}$ (where $S$ and $B$ are
the number of signal and background events) assuming an integrated
luminosity of 100fb$^{-1}$.  Table~\ref{tab:signal} shows their 
results for two models of strong EWSB:  a 1~TeV Higgs and a rescaled
$\pi\pi$ scattering model
(including chirally coupled $\rho$, $\sigma$ and $f_0$ states) scaled
from QCD.  

Finally, it was acknowledged that the existence of compositeness at
scales of order 10 TeV, as discussed in \cite{pirogov},
might give rise to new interesting resonances at
energies accessible to the VLHC.

Drawing any firm conclusions about the physics of the VLHC is currently
impossible, because we do not know what underlies electroweak symmetry
breaking.  For the purposes of this working group we assume that 
it involves some new strong dynamics. It then seems reasonable to
conclude that:
\begin{itemize}
\item The VLHC should be designed to probe
the TeV scale in detail, since the physics associated with electroweak
symmetry breaking will be there.  
This is the only scale at which we can currently say much about the
possibilities for new physics.   
If the LHC has discovered this
physics, the VLHC will be able to explore it in depth.
If this physics lies just
beyond the reach of the LHC, we will nonetheless know it must exist, and
the VLHC will catch it.  
\item 
At the same
time, the VLHC should have an ultimate reach
of 10 TeV or more at the partonic scale,
so that it can be sensitive to any
relatively low-lying phenomena associated with flavor physics.  
\item
The VLHC detector(s) will need to be capable of identifying
final states involving multiple vector bosons.  This will require good
capability for measuring the charge and momentum of leptons,
and the energy and direction of hadronic jets (including
forward jets associated with spectator
quarks).  It is not clear if missing $E_T$ measurement will have a high 
priority.  Tagging the jets associated with 
high-energy top quarks or taus would
also be very useful, because the physics of flavor-symmetry breaking might
be expected to give top- and tau-enriched signals.  Tagging heavy flavor in
the environment of the VLHC will be an interesting challenge!
\item
While a compelling case for studying the TeV scale exists,
far less is known about what might lie at higher scales.  A challenge
to theorists is to identify the possibilities for 10 TeV-scale
physics.  For instance, it would be interesting to know the extent to
which specific classes of experimental results at lower energies (e.g.
observing a particular spectrum of technipions at the LHC) would help
narrow the options at higher scales. 
\end{itemize}

\noindent
However nature has chosen to construct the world, we can be sure that
if it involves new strong dynamics the VLHC will have a rich spectrum of
new physics to explore.  

%

\clearpage

\section{Supersymmetry Working Group}

\begin{center}
{Greg W. Anderson \\ {\it Fermi National Accelerator Laboratory}} 
\end{center}

Assessing the role of a VLHC as a tool for studying supersymmetry
at this point in time is a problematic enterprise.   This question
depends largely and significantly on what is and is not seen in future 
collider experiments.
As a candidate for physics just beyond the standard model, supersymmetry
presents us with a large variety of models and frameworks.  Still, in this
context, it is reasonable to ask ``under what circumstances
would a very high energy collider like the VLHC 
further, or possibly complete, our understanding of weak-scale 
supersymmetry?''
The principal scenarios which could require a very high energy 
accelerator fall into two classes:  Models with a multi-scale 
superpartner spectrum, and models with gauge mediated supersymmetry breaking.
\footnote{Of course, other possible motivations for a VLHC, including
extended gauge groups, extended heavy Higgs sectors, or additional
heavy exotic-particles are compatible with weak-scale SUSY, 
but these augmentations are not specific to SUSY models, and these 
subjects are treated in detail by the New Strong Dynamics, and Exotics groups.}

If the physics just beyond the standard model is supersymmetric,
naturalness requires an abundance of superpartners with
masses below a few hundred GeV\cite{naturalness}.  These new spectra
should be well within the range of the LHC and a 1.5 TeV NLC, and it is 
likely the lighter of these particles will be accessible at the Tevatron
and perhaps LEP-II.  If no evidence of supersymmetry is seen by the
time the LHC is in operation, supersymmetry will have little motivation
as the physics just beyond the weak-scale.  In this case, it is supersymmetry
and not the VLHC which is lacking motivation and one would expect
new dynamics of the type discussed by other groups in the VLHC study.
If SUSY does appear at the weak-scale, it will be discovered
by the next generations of accelerators (Tevatron, LEP-II, LHC).  Moreover,
the LHC and a 1-1.5 TeV NLC could provide us with considerable information
about the spectroscopy and interactions of superparticles.
What might we learn about SUSY at these colliders that would argue for 
a higher energy machine such as the VLHC?

It is possible that pre-VLHC experiments would only uncover
part of the supersymmetric spectrum. Although the simplest formulations 
of weak-scale supersymmetry would place {\em all} of the superpartners 
below a few--several hundred GeV, it is tenable for some superpartners to 
appear at a higher scale\cite{multiscale1}.  Because 
the first two generations of squarks and sleptons 
couple less strongly to the Higgs sector, it is possible for them to have 
masses of several TeV without violating naturalness unacceptably.  
Supersymmetry requires at least two Higgs doublets, and if the multi-TeV
scale involves new dynamics, it is also conceivable for part of the Higgs 
sector to have masses as
heavy as a few TeV\cite{multiscale1}.  
  Evidence for a multi-scale superpartner spectrum would be inferred
from the absence of some modes of superpartner production.  
Moreover, the radiative
effects of multi-TeV superpartners would induce a non-decoupling violation 
of the equality between the gauge couplings of bosons and gauginos.  
This violation 
can occur at the $1-10\%$ level for multi-TeV scale superpartners
\cite{multiscale2}.

The most compelling case for a VLHC would arise if future collider 
experiments could probe the dynamics of supersymmetry breaking.
Any supersymmetric theory of physics beyond the standard model must
contain a mechanism for breaking supersymmetry and a method (messenger) for
communicating this breaking to the superpartners of the standard model
particles.  Hence, any SUSY discovery immediately implies the existence 
of two, possibly distinct, scales beyond the weak-scale:  the fundamental 
scale of SUSY breaking and the messenger scale.  A typical superpartner mass 
$\tilde{m}$ is related to the messenger scale $M$ and the dimension-two
supersymmetry breaking vev $F$ by:
\begin{equation}
\tilde{m} \propto \eta {F\over M}
\end{equation}
here we include a parameter $\eta $ as a placeholder for additional
suppressions which can be supplied by dimensionless couplings.
If supersymmetry breaking is mediated
by gravitational interactions, $M=M_{Pl}$, requiring 
$\sqrt{F} \sim 10^{10} GeV$ 
a scale so high as to be irrelevant for conceivable collider experiments.  
If supersymmetry breaking is communicated by gauge interactions,
$M$ is replaced by the mass of heavy vector-like messenger fields and
the parameter $\eta$ contains a factor $\alpha/4\pi$.  For
$\sqrt{F} \sim M$, it is possible that the messengers of supersymmetry 
breaking and the fundamental scale of SUSY breaking itself are as low
as $10-10^2$ TeV. However even with gauge mediation these relatively
low scales are not inevitable.  Anything resembling a ``no-lose''
theorem for the VLHC would rely both our ability to determine
that supersymmetry breaking has been communicated by gauge interactions,
and on our ability to place an upper bound on the messenger scale.
   
   We have at least two ways of distinguishing gauge-mediated SUSY breaking
from gravitationally mediated SUSY breaking.  In their simplest
forms, these two mechanisms for mediating SUSY breaking lead to rather 
different patterns of superpartner masses (see for example Figs. 10--11 
of Ref.~\cite{bagger}).  A more dramatic diagnostic comes from the
decays of superpartners.  In gravitationally mediated models,
the lightest superpartner--LSP,
typically the lightest neutralino -- $\tilde{\chi}^{0}_1$ is stable.
\footnote{Assuming R-parity conservation.}  In gauge
mediated models, the gravitino, with a mass typically on the 
scale of eV's  takes on the role of the LSP.
In this case, distinctive decays of the next to lightest 
superpartners--NLSP,  into the
gravitino {\it e.g.}, $\tilde{\chi}^{0}_{1} \rightarrow \gamma + \tilde{G}$
may occur inside the detector, 
leading to  signatures with two-photons, missing energy and
various combinations of jets and leptons\cite{Fayet,nlspdec1,DNS,nlspdec2}.
(For other possibilities see for example \cite{AKM}).

  The more challenging task is determining if the messenger sector
and/or the fundamental scale of supersymmetry breaking is within 
reach of the VLHC.  We can hope to learn something about a
potential multi-TeV messenger scale from the mass spectra
of superpartners and from  their decays to gravitinos.

The field(s) responsible for SUSY breaking, whether fundamental or
composite may have both supersymmetry preserving and dimension-two 
supersymmetry breaking vevs which we denote by $S$ and $F_S$
respectively.  In the simplest models, if the field responsible for
supersymmetry breaking couples to a pair of messengers
with a Yukawa coupling $\lambda$, each supermultiplet of messengers
will split into a pair of heavy and light scalars along
with an intermediate mass fermion. 
In this case messenger masses 
 can be written in terms of two parameters\cite{DNS}:
\begin{eqnarray}
M_{f} &=& \frac{\Lambda}{x} \cr
M_{\ell,h} &=& \frac{\Lambda}{x}\sqrt{1\mp x} 
\end{eqnarray}
The scale $\Lambda = F_S/S$, is roughly a factor of $10^2$ larger than a
typical superpartner mass, and will be determined once the magnitudes
of superpartner masses are measured {\it e.g.},
$m_{\tilde{g}},m_{\tilde{q}} \sim \frac{\alpha_s}{4\pi} \Lambda $.
The residual uncertainty in the values of messenger masses is
parameterized by $x = F_S/\lambda S^2 = \lambda^{-1}\Lambda^2/F_S$.  
For simplicity we neglect variations in $\lambda$ across different 
messenger representations.  In order to avoid
unwanted breaking of color, $x$ is bounded from above by 
one.  For fixed superpartner masses, as $x\rightarrow 0$, 
messenger particles become inaccessibly heavy, even for a VLHC.

The appearance of heavy messenger representations, induce
soft supersymmetry breaking masses for the SM superpartners
through loop corrections.  At a renormalization group scale
$\mu \sim M_f$  the induced gaugino masses, $M_a$ and scalar
superpartner masses $\tilde{m}$ take the form:~\cite{DNS,DGP,martin}:
\begin{eqnarray}
M_{a} &=& \frac{\alpha_a}{4\pi}\Lambda \sum_i n_a(i)g(x_i) \cr
\tilde{m}^2 &=& 2 \Lambda^2 
\sum_{a} \left(\frac{\alpha_a}{4\pi}\right)^2 C_a \sum_i n_a(i) f(x_i), 
\end{eqnarray}
where $C_a$ is the quadratic Casimir invariant of the MSSM superpartner
field, and $n_a(i)$ is the Dynkin index of the i-th messenger pair.
For the minimal ${\bf 5 + \bar{5} }$ model, the sum over messenger
representations: $\sum_i n_1(i) = \sum_i n_2(i) = \sum_i n_3(i) =1$.

The superpartner mass spectrum depends on $x$ in two ways, so 
a precise measurement of the superpartner spectrum can in principle
be used to place an upper-bound on the messenger sector.
The masses in Eq. 3 must be renormalized down to low energy.
This induces a logarithmic dependence of the superpartner 
masses on $x$.  
However, the softness of this logarithmic dependence 
makes the prospect of obtaining an upper bound on messenger masses
low enough to provide a guarantee for the discovery at a VLHC
appear quite challenging.  

In the fortunate circumstances that $x$ is close to one,
and upper-bound on the messenger scale could be achieved by
examining the ratio of gaugino and scalar superpartner masses.
For small $x$,  the functions $g(x)$ and $f(x)$ are very close
to one, and the only dependence of superpartner masses on
the messenger scale is the logarithmic dependence discussed
above.  As $x$ approaches $1$ the functions $f(x)$ and $g(x)$
depart from values close to one.  In this case examination of
scalar superpartner--gaugino mass ratios may provide a quantitative 
measure of the messenger scale.  The relevant quantity is
$\sqrt{f}/g$, which is always less than unity and approaches
$1$ for $x\rightarrow0$.
$\sqrt{f}/g \lapprox .8 \; (.9)\; ((.95))$, requires
$x \gapprox .9 \; (.72) \;((.54))$ respectively.
This approach also appears quite challenging, unless $x$
is quite close to one.  
Moreover, the simple dependence of the superpartner mass ratios
on $\sqrt{f}/g$ occurs at messenger energy scales, and this
contribution must be disentangled from renormalization effects,
the dependence on the messenger content, and
other effects. 
However, these fortuitously large
values of $x$ also coincide with the light messenger masses we have
the best chance of probing.  Our ability to use superpartner mass
measurements to place an upper bound on the messenger scale which
lie within the reach of a VLHC requires that future colliders 
make reasonably precise measurements of superpartner masses.

Recent studies of the potential for superpartner mass measurements at 
the LHC appear quite promising.   For example, 
Hinchliffe {\it et.al}, \cite{LHC}
were able to extract superpartner mass measurements at the level of
$\sim 10\% $ and $\sim 20\% $ 
for Snowmass LHC study.
However,  the precision of these measurements depends on
were one is in SUSY parameter space, and these analyses are not model 
independent.  How precisely, and model-independently we will be able
to measure superpartner masses at the LHC is not yet known.
More detailed analyses have been made concerning 
precision measurements achievable
at an NLC~\cite{nlc}.
If at all possible, bounding the messenger scale significantly
may require a future lepton collider, but whether this is a
necessity will not be clear for some time.

An upper bound on the messenger scale could also be inferred
from upper bounds on the displaced vertex in NLSP decay.
The gravitino coupling to superpartners diminishes significantly
as the scale supersymmetry breaking increases.   Accordingly,
a shorter lifetime for the NLSP requires a lower scale of supersymmetry 
breaking and lighter messenger masses.
Rewriting Eq. 2 for the messenger fermion mass,
a bound on $F$ can be translated into a bound on the messenger scale $M$:
\begin{equation}
M_f = \lambda \frac{F_S}{\Lambda}
    < \lambda \frac{F_{tot}}{\Lambda} 
      \lapprox \frac{F_{tot}}{\Lambda}
\end{equation}
here we make a distinction between $F_S$ and $F_{tot}$ because 
there may be other sources of supersymmetry breaking in addition
to the SUSY breaking field coupled to messenger fields.
Because $\Lambda$ can be in principle determined by measurements
of superpartner masses, an upper-bound on the messenger scale
can be found if we can place and upper bound on $F_{tot}$ or
equivalently an upper bound on the distance to the displaced vertex.
The decay width for  the lightest neutralino into a gravitino
in gauge mediated models is
\begin{equation}
\Gamma(\tilde{\chi}^0_1 \rightarrow \tilde{G}\gamma)
= 20 \kappa 
\left(\frac{m_{\tilde{\chi}^0_1}}{100 \; {\rm GeV}} \right)^5
\left(\frac{\sqrt{F}}{10 \,{\rm TeV}}\right)^{-4} {\rm eV}
\end{equation}
where $\kappa$ is the photino content of $\tilde{\chi}^{0}_1$.
The probability that the neutralino travels a distance $x$ before
decaying in the detector is $P(x) = 1 - e^{-x/L}$, where
\begin{eqnarray}
L =  \frac{ 9.9 \times 10^{-3} \mu m} {\kappa}
\left( \frac{ m_{\tilde{\chi}^0_1}}{100 GeV} \right)^{-5} 
\left( \frac{\sqrt{F_{tot}}}{10\;{\rm TeV}} \right)^{4}  \nonumber \\ 
\left(E^{2}_{\tilde{\chi}^0_1}/m^{2}_{\tilde{\chi}^0_1} -1 \right)^{\frac{1}{2}}.
\end{eqnarray}
Leading to the bound:
\begin{eqnarray}
M_f \lapprox 10 \; {\rm TeV}
\left(\frac{100 \; {\rm TeV}}{\Lambda} \right)
\left(\frac{\kappa L}{0.99 \; \mu{\rm m}} \right)^{1/2} \nonumber \\
\left(\frac{100\;{\rm GeV}}{m_{\tilde{\chi}^0_1}} \right)^{5/2} 
\left(E^{2}_{\tilde{\chi}^0_1}/m^{2}_{\tilde{\chi}^0_1} -1 \right)^{-\frac{1}{4}}
\end{eqnarray}
So this method will challenge our ability to resolve 
relatively small displaced vertices as well.

  In both cases, establishing that messengers lie within reach
of a VLHC requires relatively large values of $x$, relatively light values
of $F$, and reliable measurements of superpartner masses.
On one hand combining these fortunate circumstances
may appear to be wishful, but the fortunate circumstances under which
we would be able to identify and place reliable upper-bounds messenger 
scale overlap with those where multi-TeV messenger states are light enough 
to be accessible at a VLHC.  Good luck is when preparation and 
opportunity meet, and we should be prepared to exploit this opportunity
if it arises.

We conclude with a few remarks about the collider signatures
of the messenger sector.  If light enough, messenger particles
would be pair produced at the VLHC.
Heavy messenger scalars will decay to messenger fermions by radiating
gauginos and messenger fermions will decay to the lighter
messenger scalars by radiating gauginos as well.  
Renormalizable Yukawa interactions between messenger fields
and standard model fields potentially introduce flavor changing
neutral currents, spoiling  the principle motivation for 
low energy $10$--$100$ TeV SUSY breaking.  In the absence of such
couplings the lightest messenger fields contain conserved quantum
numbers and are stable.  The presence of nonrenormalizable operators
may induce messenger decay, but not on time scales relevant to
collider searches.  Planck mass suppressed dim-5 operators, would
for example lead messengers to decay with lifetimes of 
$\sim 10^{-1}$--$10^{-2}\,s$\cite{DGP}.

The apparent unification of gauge couplings at high energies
$\sim 10^{16}$ GeV is most naturally accommodated by messenger 
representations in complete $SU(5)$ representation.
In the minimal messenger model, the messengers are contained in
the ${\bf 5 + \bar{5} }$ representation of $SU(5)$.  Under the
standard model gauge group $SU(3)\times SU(2)\times U(1)$. The
${\bf \bar{5}}$ representation decomposes as:
\begin{equation}
{\bf \bar{5}  = (\bar{3},1,\frac{1}{3}) + (1,2,-\frac{1}{2})}.
\end{equation}
Together with the $5$, the lightest scalar messenger states
will these states will have the quantum
numbers of
and has the same quantum numbers as $SU(2)$-singlet 
down squarks-- $D^c$, 
and left handed slepton doublet -- $L$ respectively.
The colored scalar messenger states hadronize to form 
multi-TeV objects with the same quantum numbers as a neutron or
proton, and would look like a canon ball in the detector.

In the absence of a discovery of supersymmetry at or
before the LHC SUSY provides little motivation for a VLHC.
However, it should be understood in this case that it is
supersymmetry and not the VLHC which is lacking motivation.

If the world is supersymmetric above the weak scale and supersymmetry 
breaking is transmitted to the standard model superpartners by gauge
interactions, the VLHC may be a logical step in the worlds future high energy
physics program.  However, this is not inevitable.
Supersymmetry could be found at the several hundred GeV scale without 
giving us any compelling reason to expect another layer of structure
at the multi-TeV scale.  The case for such a machine will rest
on  what nature provides for us, and on our ability to exploit the 
Tevatron, LEP-II and the LHC.

\clearpage
\section{Exotics Working Group}

\begin{center}
{U.~Baur \\ {\it SUNY at Buffalo} \\
S.~Eno \\ {\it University of Maryland}}
\end{center}

\subsection{Introduction}
We summarize the reach of the VLHC for contact interactions and new heavy
particles in non-supersymmetric extensions of the Standard Model. 

The Standard Model (SM) of strong and electroweak interactions, based on the
gauge group $SU(3)_C\times SU(2)_L\times U(1)_Y$, has been extremely
successful phenomenologically. It has provided the theoretical
framework for the description of a very rich phenomenology spanning 
a wide range of energies, from the atomic scale up to the $Z$ boson 
mass. However, the SM has a number of shortcomings. In 
particular, it does not explain the origin of mass, the observed
hierarchical pattern of fermion masses, and why there are three
generations of quarks and leptons. It is widely believed that at high 
energies deviations from the SM will appear, signaling the presence of 
new physics. 

Many theoretical models which attempt to overcome the shortcomings of
the SM either involve new gauge symmetries, or predict that quarks and
leptons are composite objects. A common feature of these models are new 
interactions and new
heavy particles. The mass of these objects is in general given by the
energy scale of the new interaction. At low energies, their existence is
signalled by four fermion contact interactions. A hadron collider with 
a center of mass energy of 100~TeV or more (VLHC) would offer an 
excellent chance to search for contact interactions and also, directly,
for the new heavy particles associated with new interactions.

In this brief report, we discuss the potential of the VLHC to search for
contact interactions associated with quark and lepton compositeness, and 
illustrate the discovery mass reach
for new heavy states by describing the search for excited 
quarks~\cite{harris} in some detail. In addition, we list benchmark 
results for additional gauge bosons~\cite{rizzo1} and 
leptoquarks~\cite{rizzo2},
which are predicted by many grand unified models. We also briefly 
comment on the discovery mass reach for colorons~\cite{liz} and
axigluons~\cite{carl} which appear in
models with extended strong interaction gauge symmetries. Supersymmetric
and technicolor particle searches at the VLHC are described in 
Refs.~\cite{susy} and~\cite{techni} and are therefore not discussed
here.

\subsection{Contact Interactions}

The repetition of the three generations of quarks and leptons suggests
that they are bound states of more fundamental fermions, and perhaps
bosons, bound together by a new interaction which is characterized by an
energy scale $\Lambda^+$. At energies much smaller than $\Lambda^+$, the
substructure of quarks and leptons is signalled by the appearance of
four fermion contact interactions which arise from the exchange of bound
states of the subconstituents~\cite{elp}. The lowest order contact terms are
dimension~6 four-fermion interactions which can affect jet and Drell-Yan
production at a hadron collider. Compared with the SM terms, they are
suppressed by a factor $1/\Lambda^{+2}$. The signature for four-quark contact
interactions, for example, would be an excess of events at large
transverse energy, $E_T$, similar to that observed by CDF in inclusive
jet production at the Tevatron in 
Run~1a~\cite{cdfet}. 

\begin{figure}[t]
\leavevmode
\begin{center}
\resizebox{8.9cm}{!}{%
\includegraphics{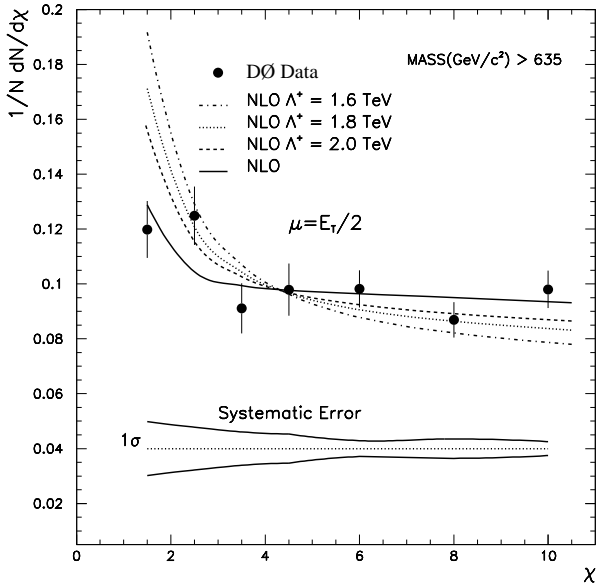}}
\end{center}
\caption{The D\O\ dijet angular distribution (points) for dijet masses
larger than 635~GeV compared to
predictions of NLO QCD (solid line), and NLO QCD with a quark contact
interaction for various values of $\Lambda^+$.}
\label{FIG:ONE}
\end{figure}

However, from the CDF measurement of the jet inclusive cross section it
is apparent that it is difficult to discover a signal for contact
interactions by looking for an excess of events at high transverse
energies, due to uncertainties in the parton distribution functions,
ambiguities in QCD calculations, and systematic uncertainties in jet
energy measurements~\cite{joey}. Another signal for quark -- quark
contact interactions, which is not very sensitive to theoretical or jet
energy uncertainties, is the dijet angular distribution which is more
isotropic than that predicted by QCD if contact terms are present. Both
CDF~\cite{tev} and D\O~\cite{bhat} have found good agreement with the 
shape predicted by QCD. The D\O\ dijet angular distribution is shown in
Fig.~\ref{FIG:ONE}. The quantity $\chi$ shown here is related to the 
scattering angle in the center of mass frame, $\theta^*$, by
$\chi=(1+|\cos\theta^*|)/(1-|\cos\theta^*|)$.
Using a model with left-handed contact
interactions, D\O\ sets a preliminary 95\% confidence level (CL) limit on
the interaction scale, $\Lambda^+$, of $\Lambda^+>2.0$~TeV~\cite{bhat}.
CDF obtains a 95\% CL limit of $\Lambda^+>1.8$~TeV~\cite{tev}.
From the inclusive
jet analysis, the dijet angular distribution and other searches 
for contact interactions at the
Tevatron~\cite{tev1}, as well as searches at the CERN $p\bar p$ 
collider~\cite{ua2} and simulations carried out for the 
LHC~\cite{reach}, we conclude that the compositeness scale reach of 
a hadron collider is roughly equal to its center of mass energy,
$\sqrt{s}$. Detailed simulations for the VLHC, however, have not been
carried out so far.

\subsection{Excited Quarks}

Conclusive evidence for a new layer of
substructure would be provided by the direct observation of excited
states of the known quarks and leptons. In the following we shall
concentrate on excited quarks with spin~1/2 and weak isospin~1/2. The
coupling between excited spin~1/2 quarks, ordinary quarks and gauge
bosons is uniquely fixed to be of magnetic moment type by gauge
invariance. Excited quarks decay into quarks and a gluon, photon or a
$W/Z$ boson, or, via contact interactions into $\bar qqq'$ final
states~\cite{BSZ}. Subsequently, only decays via gauge interactions
are considered. Excited quarks are then expected to decay predominantly
via strong interactions; radiative decays and decays into a quark and a
$W/Z$ boson will typically appear at ${\cal O}(\alpha/\alpha_s)$, {\it
i.e.} at the few per cent level~\cite{BHZ}.

\begin{figure}[t]
\leavevmode
\begin{center}
\resizebox{8.9cm}{!}{%
\includegraphics{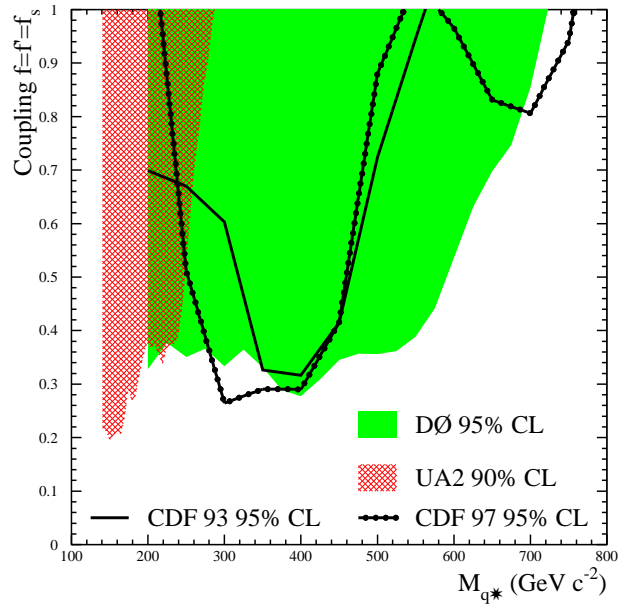}}
\end{center}
\caption{The region of the excited quark coupling versus mass plane
excluded by UA2, CDF and D\O\ measurements.}
\label{FIG:TWO}
\end{figure}

In hadronic collisions, excited quarks can be produced singly via quark
gluon fusion. The subsequent $q^*\to qg$ decay leads to a peak in the
two jet invariant mass distribution located at $m(jj)=M^*$, where $M^*$
is the excited quark mass. UA2~\cite{qstarua2}, 
CDF~\cite{cdfqstar,cdfqstar2} and D\O~\cite{d0qstar} have searched for 
$q^*$ production in the
dijet invariant mass distribution. Figure~\ref{FIG:TWO} shows the region
of the excited quark coupling $f=f'=f_s$ versus $M^*$ plane excluded by
those experiments. Here, $f$, $f'$ and $f_s$ are the strength of the
$SU(2)_L$, $U(1)_Y$ and $SU(3)_C$ couplings of the $q^*$ to quarks and
the SM gauge fields when the scale of the magnetic moment coupling
is set equal to $M^*$.
For $f=f'=f_s=1$, CDF sets a lower 95\% CL limit of $M^*>760$~GeV 
(with the exception of the region $570~{\rm GeV}<M^*<580$~GeV), whereas
D\O\ finds a (preliminary) bound of $M^*>725$~GeV (95\% CL). Only first
generation excited quarks, $u^*$ and $d^*$, are considered. The $u^*$
and $d^*$ are assumed to be degenerate in mass. 

The discovery reach of the VLHC in this model has been studied in 
Ref.~\cite{harris}, assuming a Gaussian dijet invariant mass resolution
of $\sigma=0.1\,m(jj)$, which is similar to that of the CDF detector. Since
$\Gamma(q^*)\approx 0.04\,M^*$ in the model considered, approximately
90\% of the two jet events from an excited quark will be in the mass window
$0.84\,M^*<m(jj)<1.16\,M^*$. To estimate the mass reach, the
differential cross section is integrated within this window for both the
$q^*$ signal and the QCD background. The QCD background rate is then
used to find the $5\,\sigma$ discovery cross section. This is defined as
the cross section which is above the background by $5\,\sigma$, where
$\sigma$ is the statistical error on the measured cross section.

\begin{figure}[t]
\leavevmode
\begin{center}
\resizebox{8.9cm}{!}{%
\includegraphics{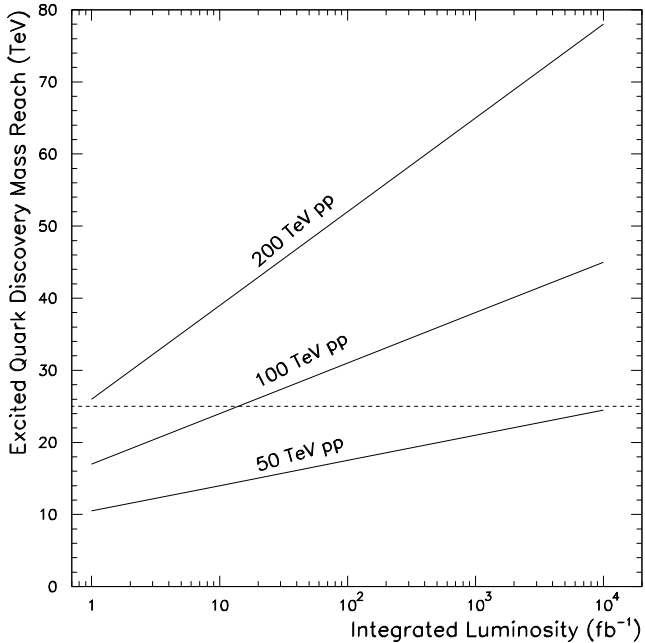}}
\end{center}
\caption{The $5\,\sigma$ discovery mass reach for $pp\to q^*\to jj$ is
shown as a function of the integrated luminosity for a VLHC with center 
of mass energy of 50~TeV, 100~TeV and 200~TeV (solid lines). The horizontal
dashed line illustrates what integrated luminosity is necessary to
discover an excited quark with mass $M^*=25$~TeV. }
\label{FIG:THREE}
\end{figure}

The discovery mass reach for excited quarks at the VLHC is shown in 
Fig.~\ref{FIG:THREE}
for three different machine energies as a function of the integrated
luminosity.
For an integrated luminosity of $10^4~{\rm fb}^{-1}$, the mass reach 
at a $pp$ collider with center of mass energy of 50~TeV (200~TeV) is 
$M^*=25$~TeV (78~TeV). However, an excited quark with a mass of 25~TeV
would be discovered at a $pp$ collider with $\sqrt{s}=100$~TeV with
only $13~{\rm fb}^{-1}$. In this case, doubling the collider energy is
equivalent to an increase in integrated luminosity of almost a
factor~1000. A similar result is obtained from other heavy particle
searches. 

\subsection{Additional Vector Bosons and Leptoquarks}

The discovery of new gauge bosons, $W'$, $Z'$, would signal an extension
of the SM gauge group by an additional factor such as $U(1)$ or $SU(2)$.
$Z'$ bosons appear in most Grand Unified Theories. $W'$ bosons
are typical for models which restore the left-right symmetry at high
energies. 
The mass reach of a hadron collider for new gauge bosons is model
dependent due to the variations in their couplings to quarks and
leptons. At hadron colliders, new gauge bosons can be produced directly
via quark -- antiquark annihilation, $q\bar q'\to W'$ and $q\bar q\to
Z'$. CDF~\cite{cdfwpr} and D\O~\cite{d0wpr} have searched for 
$W'$ (including righthanded $W$ bosons) and $Z'$ bosons in a variety of 
models. The limits obtained vary between 565~GeV and 720~GeV. 
The discovery reach of the VLHC for new gauge bosons has been
investigated in Ref.~\cite{rizzo1}. Only the leptonic decays of the $W'$
and $Z'$ bosons, which are virtually background free, were used in this 
analysis. At a 200~TeV $pp$ collider, with an integrated luminosity of
1000~fb$^{-1}$, $Z'$ bosons with mass up to $M_{Z'}=40$ --~50~TeV can be
detected~\cite{rizzo1}, whereas the mass reach for $W'$ bosons is 
50 --~60~TeV, depending on the details of the model considered. 

Many Grand Unified Theories predict the existence of leptoquarks, $LQ$, 
which are spin~0 (scalar) or spin~1 (vector) color triplet objects 
coupling to a quark --
lepton pair. Searches for leptoquarks have been performed at
LEP~\cite{lqlep}, HERA~\cite{lqhera} and the Tevatron~\cite{lqtev}. The
most stringent bounds presently come from Tevatron data which exclude,
at 95\% CL, 
scalar (vector) first generation leptoquarks with $B(LQ\to eq)=\beta=0.5$ if
their mass is $M_{LQ}<192$~GeV (270~GeV)~\cite{norman}. Leptoquarks can
be produced either singly or in pairs at a hadron collider. The cross
section for single production depends on unknown Yukawa couplings of the
leptoquark. Pair production, on the other hand, proceeds via QCD
interactions, and therefore depends only on the leptoquark spin.
Unless the Yukawa couplings are large, pair production dominates. The
discovery reach of the VLHC for leptoquarks has been studied in
Ref.~\cite{rizzo2}. At a proton -- proton collider with
$\sqrt{s}=200$~TeV and an integrated luminosity of 1000~fb$^{-1}$, one
should be able to detect scalar (vector) leptoquarks with a mass up to
$M_{LQ}\approx 14$~TeV ($M_{LQ}\approx 23$~TeV), if $\beta=0.5$. For
$\beta=1$, the mass reach is higher by about 1~TeV.

If leptoquarks are sufficiently light, the VLHC will make it possible to
probe the Yukawa couplings of leptoquarks. At a 100~TeV proton -- proton
collider, Yukawa couplings as small as $\lambda/e= 10^{-3}$ ($\lambda/e=
10^{-5}$) [$e$ is the electric charge unit] can be probed for
$M_{LQ}=2$~TeV ($M_{LQ}=1$~TeV)~\cite{tom}.

At a $p\bar p$ collider, the search limits for new gauge bosons and
leptoquarks are both about 20 -- 30\% higher than those found for $pp$
collisions with the same center of mass energy~\cite{rizzo1,rizzo2}.

\subsection{Colorons and Axigluons}

In the coloron~\cite{liz} model, the symmetry group of the strong
interactions, $SU(3)_C$, is replaced by a $SU(3)_1\times SU(3)_2$ 
group, which is spontaneously broken to $SU(3)_C$ at an unknown
scale $v$. The corresponding color octet of massive gauge
bosons (colorons) couples vector-like to quarks. In the 
axigluon~\cite{carl} model, the gauge group is $SU(3)_L\times SU(3)_R$,
and the coupling of the massive color octet vector bosons (axigluons) 
to quarks is axial vector-like. Colorons and axigluons can
be produced via $q\bar q$ annihilation, and lead to a peak in the two
jet invariant mass distribution, very much like an excited quark. CDF
has searched for these particles in the dijet channel, and 
places a lower limit of 980~GeV (95\% CL) on their mass~\cite{cdfqstar}.
The discovery reach of the VLHC for these particles has not been
estimated yet. It is expected that colorons and axigluons in the 
multi-ten TeV range can be discovered at a 200~TeV $pp$ collider. 

\subsection{Conclusions}

In this brief report, we have discussed the search for contact
interactions and new heavy particles, which appear in popular
non-supersymmetric extensions of the SM, at the VLHC.
The search potential of the VLHC for these new
states is truly enormous; for a collider with a center of mass energy of
100~TeV or more, the limits are in general in the multi-ten TeV region.
To maximize the heavy particle search potential, the VLHC should strive
to the highest energy possible.
It should be emphasized, however, that there are no models which firmly 
predict the existence of new particles in the region of interest for
the VLHC. On the other hand, in a
situation where first signs of new physics are observed at the LHC in 
the form of contact interactions, but the scale of new
physics is too high to allow production of the associated new states 
directly, the VLHC will be a perfect tool for an in-depth investigation
of the beyond the Standard Model frontier.

\subsection{Acknowledgments}
We would like to thank H.~Frisch for stimulating discussions.
This research has been supported in part by the National Science
Foundation, grant PHY-9600770, and the Department of Energy, grant
DEFG0296ER41015.

%

\clearpage

\section{Full Rapidity Physics Working Group }

\begin{center}

{Andrew Brandt\\
{\it Fermi National Accelerator Laboratory}\\
Cyrus Taylor\\
{\it Case Western Reserve University}\\}

\end{center}

Members of the
working group include: 
M. Albrow, W. Baker, J. Bantley, P. Bloch, A. Brandt,
K. Goulianos, P. Hanlet, T. Heuring, C. Lundstedt,
H. de Motta, F. Paige, D. Rainwater,
V. Simak, A. Santoro, G. Snow, C. Taylor, and L. Voyvodic.

Particle production at a VLHC operating at $\sqrt{s}=100$ TeV 
will span some 24 units of rapidity.
Such an accelerator should  
include a detector and interaction region
optimized for full acceptance\cite{Snowmass1}.  
Design goals for such a detector
should include:
\begin{itemize}
\item all charged particles, photons and neutrons of generic
$p_t$ should be observed and their energies/momenta well measured over
all of phase space
\item diffractive and elastically scattered protons should be well
measured
\item muon identification should be extended into the far forward
regions
\item the physics of rapidity gaps should not be compromised
\end{itemize}

While no full acceptance detector has ever operated at collider
energies, such a detector, FELIX, is being proposed for the
LHC\cite{FELIX}. The lessons learned in the design
and operation of FELIX will provide the basis for a full acceptance
detector at the VLHC.

The need for a full acceptance detector at the VLHC follows from 
basic kinematics. Physics on the energy
frontier is necessarily central and will largely be the domain
of optimized central detectors, conversely, any physics
not on the energy frontier is forward physics, and will benefit from
a full acceptance detector.

A second point is that a full acceptance detector should operate in
an environment of $\sim 1$ interaction per beam crossing.  This 
ensures that global event structure can be determined, event by
event. In contrast, central detectors operating on the energy frontier 
will also be operating on the luminosity frontier, with many collisions
per beam crossing.  This observation also has a corollary:  any precision
physics or standard physics will likely be done well by a full acceptance
detector.

Finally, one should state the obvious:  a central detector, optimized for
high-$p_t$ physics at the energy frontier in messy environments will only
be sensitive to a small fraction of the kinematically allowed phase space.
The discovery potential of a full acceptance detector operating in 
unexplored regions of phase space should thus be noted.

A few examples  illustrate the scope of a full acceptance detector
at the VLHC. 

\subsection{The small-x frontier}

Hard processes at the VLHC span a kinematically allowed region in
$(x,Q^2)$ given by
\begin{equation}
x_1 x_2 \geq {4 E_t^2\over s},
Q^2 \geq E_{tmin}^2
\end{equation}
where $x_1,x_2$ are the momentum fractions of the two partons
involved in the hard scattering, and $E_{tmin}$ is the minimum 
transverse energy needed to identify the process.  Thus, at
$\sqrt{s}= 100$ TeV, for scattering to two jets with 
$E_{tmin}\sim 10$ GeV, one can probe the proton structure down to
$x\sim 4*10^{-8}$.  This is some 4 orders of magnitude smaller
than the HERA limit, and will be an extremely interesting domain
of QCD\cite{asw}.

\subsection{Forward particle tags}

Particle production in the fragmentation region has never been
studied at collider energies. A full acceptance detector, with
complete coverage for neutral particles down to zero degrees, and
with complete charged particle tracking, will  not only measure
such production exquisitely, it will also be able to tag leading
particles.  For example, detecting leading deltas or neutrons
tags the rest of the event as a collision between a beam nucleon
with the exchanged non-strange meson.  The VLHC thus becomes
an effective meson-proton collider.  One can similarly tag on both
sides, defining meson-meson interactions, tag on strange meson
exchange, and so on.  Aside from the rich physics program that
this capability will allow, it is also necessary for total cross
section measurements which are crucial for determining the cross
sections for all physics processes.
 
\subsection{Rapidity gap phenomena}

While perturbative QCD has been extremely successful at describing
and predicting many aspects of the strong interactions, many fundamental
processes cannot yet be calculated in this language.  Processes such
as elastic and diffractive scattering are instead still understood
in terms of Regge theory, with elusive objects like the ``Pomeron''
playing a central role in current phenomenology.  The study of such
processes, in particular hard diffractive processes, has expanded
dramatically in recent years\cite{Snowmass2}, with pioneering work
at UA8, followed by important ongoing studies at HERA and 
the Tevatron.
It should be noted that all of these current or past experiments would
have benefited tremendously from increased coverage.  A full acceptance
detector at the VLHC will be able to do all of this physics superbly,
allowing a continuous transition from the clearly perturbative
regime into the non-perturbative regime in a controlled manner.  

\subsection{Cosmic ray phenomena}

All experimental information about particle interactions at the 
highest energies comes from cosmic ray experiments.  LHC energies
correspond
to primary cosmic rays of about 100 PeV ($10^{17}$ eV); particles with
energies of $10^{20}$ eV have been observed.  While fraught with problems
of limited statistics and complicated systematics, cosmic ray
experimentalists
can point with pride to a number of important discoveries, including
a pre-discovery of charm.  It is thus important to take note of the fact
that there are a number of anomalies reported in studies of cosmic ray
interactions hinting at unusual physics, not anticipated in the standard
model.  These anomalies are observed at energies beyond the reach of 
current accelerators.  Further, since cosmic ray experiments track
energy flow, their sensitivity is typically in the fragmentation region,
beyond the reach of current (central) collider detectors. A full
acceptance
detector should be designed keeping these anomalies in mind.  The list
of anomalies includes reports of anomalous mean free paths, anomalous
forward heavy flavor production, anomalous attenuation of secondary
hadrons, anomalies in the energy fraction of air showers, scaling
anomalies, anomalies in the charged to neutral ratio (Centauros and
anti-Centauros), and anomalously parallel multi-muon bundles.  Only
detectors with good acceptance in the very forward direction can 
test these claims in an accelerator environment.



\subsection{Summary}
 
The task of designing a full acceptance detector for the VLHC is
non-trivial, and requires careful coordination with the design of
the machine itself.  The starting point is necessarily the magnetic
architecture, which must be integrated into the machine lattice.
The design of FELIX, a possible full
acceptance detector for the LHC, should serve as a prototype 
for discussing
the design of a full rapidity detector at the VLHC.

Important features of the FELIX design which may translate to
the VLHC include:
\begin{enumerate}
\item The relatively low luminosity should permit an insertion
in which focusing quadrupoles are located
a large distance from the
collision point.  In FELIX, this distance is more than 100 m.
\item The requirement of complete calorimetric coverage for
neutrals demands a precision zero degree calorimeter. This requires
that the beams be separated by a significant transverse distance
(in FELIX, 42 cm) at the location of the zero degree calorimeter.
The beam separation is defined in FELIX by the requirement that
the experiment co-habit with the RF cavities, located 140 m from
the beam.
\item The dipole fields needed to move the beams through the
experimental area within the above constraints will play a dual
role as spectrometer analysis magnets and consequently
should have the 
largest possible aperture.
\item While the central region is not the focus of the proposed
experiment, it should nevertheless have a good central detector.
Such a detector might be built around elements of the preceeding
generation of collider detectors.  It need not be of the quality
of the high $p_t$ central VLHC detector, but should also not be
neglected.
\end{enumerate}

Although it is premature to begin
detailed work on possible optics for an insertion
at the VLHC, the need for a long straight section is clear and 
should be built into any VLHC design at the earliest stage.
Simple scaling with the
beam energy (which seems reasonable given constants such as transverse 
shower sizes in calorimeters) would suggest that the zero degree
calorimeter should be located at least 700 m from the collision
point.  The required beam separation at this point would imply 
that a straight section with a total
length of 2.8--4 km seems appropriate.

A full acceptance detector will provide a powerful
tool for the study of physics at the VLHC.  While low in cost compared
to detectors concentrating on physics at the energy frontier in the
central region, a full acceptance detector will combine a strong program
of physics complementary to other detectors with a substantial 
discovery potential, particularly for the ``unexpected''.

\clearpage

\section{Precision Measurements Of Heavy Objects Working Group}

\begin{center}

{Frederick Olness\\
{\it Southern Methodist University}\\
Randall Scalise\\
{\it Southern Methodist University}\\}

\end{center}

\subsection{Introduction} 
We report on the activities of the 
Precision Measurements of Heavy Objects working group.
The following people contributed to the writing of this summary:
Marcel Demarteau, Vassilis Koulovassilopoulos,
Joseph Lykken, Stephen Parke(convener during the workshop),
Erich Varnes, G. P. Yeh (convener during the workshop).

The topics discussed by the  Precision Measurements of Heavy Objects
working group spanned a very wide range; consequently, it is impossible
to cover each topic in depth. 
 Therefore,  in this report we will primarily focus
on the issues most relevant to a VLHC machine. 
 In the following, we mention only the highlights, and
refer the reader to the literature for more specific questions.

\subsection{Parton Distributions for VLHC}


\def\figFredi{
\begin{figure}[htbp]
\begin{center}
\leavevmode
 \epsfxsize=3in  \epsfbox{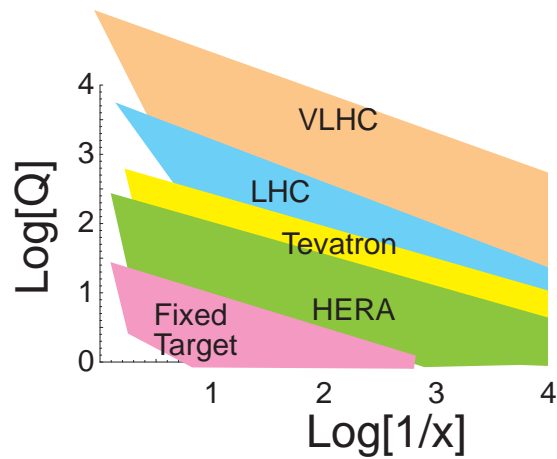}
\end{center}
\caption{Kinematic range of various machines.
Note the small $x$ range is clipped in this plot.  The $Q$ scale is in
GeV and the logs are base 10.
}
\label{fig:Fredi}
\end{figure}
}


\def\figFredii{
\begin{figure}[htbp]
\begin{center}
\leavevmode
 \epsfxsize=1.7in  \epsfbox{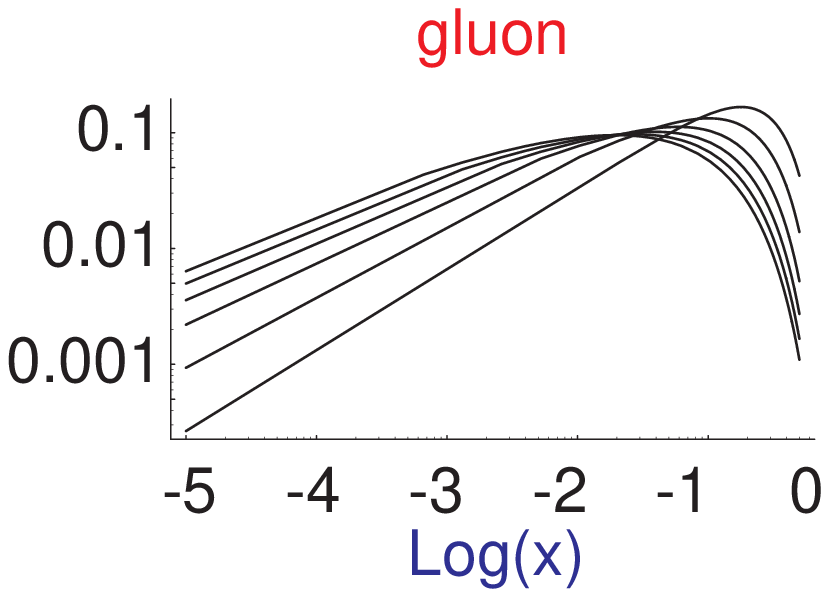}
 \epsfxsize=1.7in  \epsfbox{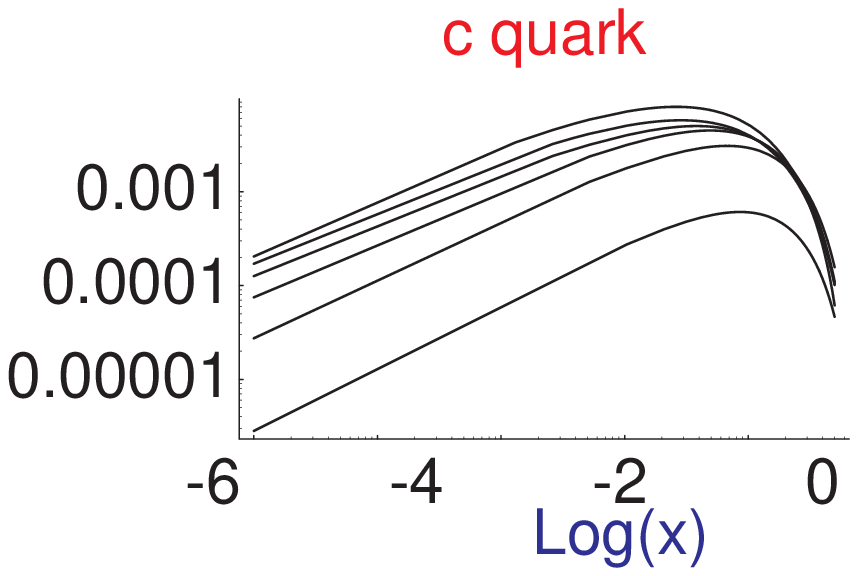}
\end{center}
\caption{Evolution of the a) gluon and b) charm PDF's
in $Q$ {\it vs.} $x$.
We display $x^2 f_{i/P}(x,Q)$ for 
$Q=\{2, 10^1,10^2,10^3,10^4,10^5 \}$ GeV.
}
\label{fig:Fredii}
\end{figure}
}


\def\figFrediii{
\begin{figure}[htbp]
\begin{center}
\leavevmode
 \epsfxsize=1.7in  \epsfbox{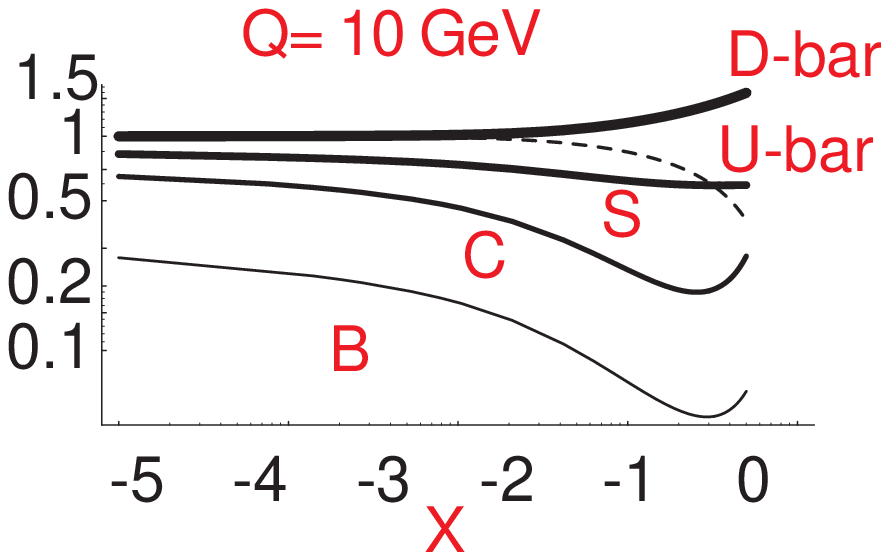}
 \epsfxsize=1.7in  \epsfbox{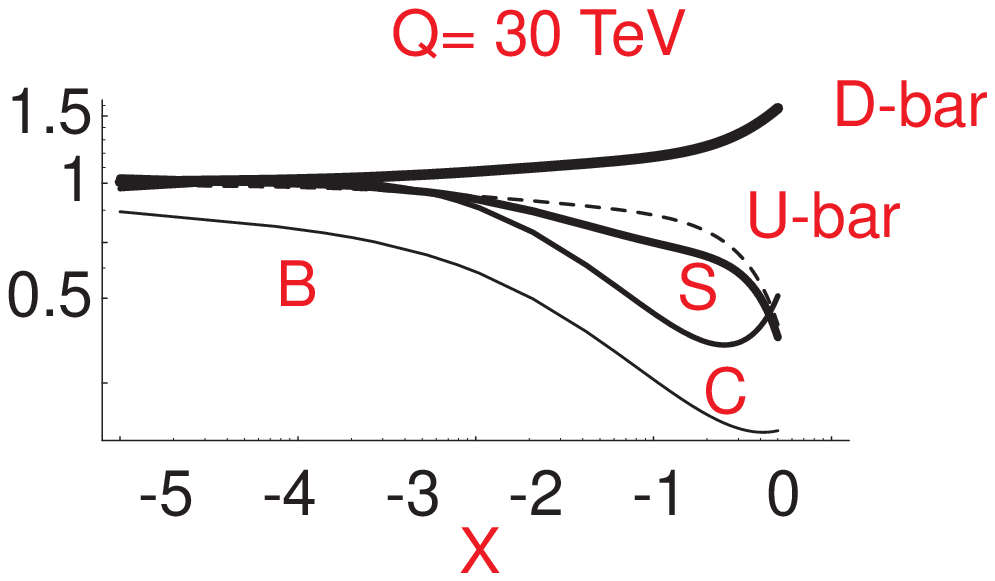}
\end{center}
\caption{Flavor democracy at a) 10 GeV and b) 30 TeV.
We compare the individual parton distributions $f_{i/P}(x,Q)$
to that of the average sea, $(\bar{u} + \bar{d})/2$.
}
\label{fig:Frediii}
\end{figure}
}

Global QCD analysis of lepton-hadron and hadron-hadron  processes has
made steady progress in testing the consistency of perturbative QCD
(pQCD) within many different sets of data, and in yielding
increasingly detailed information on the universal parton
 distributions.\footnote{PDF sets are available via WWW on the CTEQ page at
http://www.phys.psu.edu/$\sim$cteq/ and  on the The Durham/RAL HEP
Database at  http://durpdg.dur.ac.uk/HEPDATA/HEPDATA.html.}

 We present  the kinematic ranges covered by selected facilities
relevant for the determination of the universal parton distributions.
 While we would of course like to probe the full $\{x,Q\}$ space,
the small $x$ region is of
special interest.  For example, the rapid rise of the $F_2$ structure
function observed at HERA  suggests that we may reach the parton
density saturation region more quickly than anticipated.
Additionally, the small $x$ region can serve as a useful testing
ground for BFKL, diffractive phenomena, and similar processes.
Conversely, the production of new and exotic phenomena generally
happens in the region of relatively high $x$ and $Q$.

 This  compilation provides a useful guide to the planning of future
experiments and to the design of strategies for global analyses.
Another presentation regarding future and near-future machines
is given in the 1996 Snowmass Structure Functions Working Group 
report \cite{sfun}.

Here we will simply mention a few features which are particularly relevant
for such a very high energy facility as a VLHC.

\figFredi

As we see in Fig.~\ref{fig:Fredi}, the VLHC will probe an $\{ x,Q\}$ region 
far beyond the range of present data.  To accurately calculate processes
at a VLHC, we must have precise PDF's in this complete kinematic
range.  Determining the PDF's in the small $x$ regime is a serious
problem since there will be no other measurement in the extreme
kinematic domain required by VLHC.  For the large $x$ and $Q$ region, the
PDF's at large $Q$ can, in principle, be determined via the standard
QCD DGLAP evolution, but in practice uncertainties from the small $x$
region can contaminate this region.

\figFredii

\figFrediii

In Fig.~\ref{fig:Fredii}, we display the evolution of the PDF's for a 
selection of 
partons.  For the gluon and the valence quarks, we see a decrease at high $x$
and an increase at low $x$ with $x \sim 0.1$ as the crossing point.
In contrast, for the heavy quark PDF's, we see generally an increase 
with increasing $Q$.  The momentum
fraction of the partons {\it vs.} energy scale is shown in 
Table~\ref{tab:momfrac}.
An interesting feature to note here is the approximate ``flavor
democracy" at large energy scales; that is, as we probe the proton at
very high energies, the influence of the quark masses becomes
smaller, and all the partonic degrees of freedom carry comparable
momentum fractions.  To be more precise, we see that at the very
highest energy scales relevant for the VLHC, the strange and charm
quark are on par with the up and down sea, (while the bottom quark
lags behind a bit).  This feature is also displayed in Fig.~\ref{fig:Frediii}
where we show these contributions for two separate scales.  In light of this
observation, we must dispense with preconceived notions of what are
``traditionally" heavy and light quarks, and be prepared to deal with
all quark on an equal footing at a VLHC facility. This approach is
discussed in the following section.

\def\bigstrut{\vrule height16pt depth6pt width0pt}%
\begin{table}[htbp]
\begin{center}
\caption{Momentum fraction (in percent) carried by separate partons as a 
function of the energy scale $Q$.
}
\label{tab:momfrac}
\begin{tabular}{||r|c|c|c|c|c|c||} \hline \hline 
\bigstrut
Q \qquad & 	$g$	 & $\bar{u}$	 & $\bar{d}$	 & $s$	 & $c$	 
& $b$  \\ \hline 
  3 GeV & 46 & 5 & 7 & 3 & 1 & 0  \\ \hline 
 10 GeV & 48 & 6 & 8 & 4 & 2 & 0  \\ \hline 
 30 GeV & 48 & 6 & 8 & 5 & 3 & 1  \\ \hline 
100 GeV & 48 & 7 & 8 & 5 & 3 & 2  \\ \hline 
300 GeV & 49 & 7 & 8 & 6 & 4 & 2  \\ \hline 
  1 TeV & 49 & 7 & 8 & 6 & 4 & 3  \\ \hline 
  3 TeV & 49 & 7 & 8 & 6 & 4 & 3  \\ \hline 
 10 TeV & 50 & 7 & 9 & 6 & 5 & 4  \\ \hline 
 30 TeV & 50 & 7 & 9 & 7 & 6 & 4  \\ \hline 
100 TeV & 51 & 7 &10 & 7 & 7 & 4  \\ \hline  
\hline 
\end{tabular}
\end{center}
\end{table}

\subsection{Heavy Quark Hadroproduction}


\def\figHQdata{
\begin{figure}[htbp]
\begin{center}
\leavevmode
 \epsfxsize=3in  \epsfbox{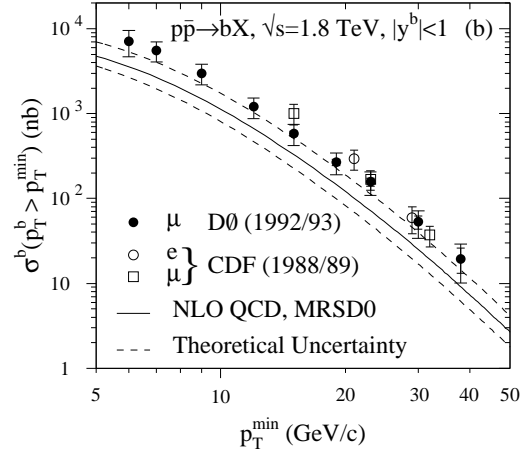}
\end{center}
      \caption{
Heavy quark hadroproduction data. 
{\it Cf.}, Ref.~\protect\cite{hqdata}.
}
   \label{fig:figHQdata}
\end{figure}
}

\def\figProd{
\begin{figure}[htbp]
\begin{center}
\leavevmode
\hbox{
 \epsfxsize=0.40\textwidth  \epsfbox{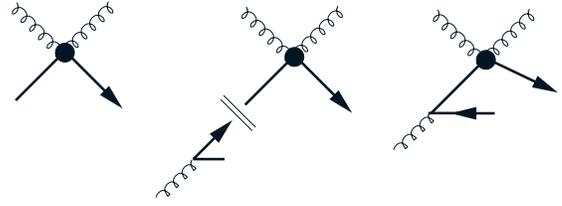}
}
\end{center}
      \caption{
a)~Generic leading-order diagram for heavy-flavor excitation (LO-HE), $gQ\to gQ$.
b)~Subtraction diagram for heavy-flavor excitation (SUB-HE),
   ${}^1f_{g\to Q} \otimes \sigma(gQ\to gQ)$.
c)~Next-to-leading-order diagram for heavy-flavor creation (NLO-FC).
\null\hfill\null}
   \label{fig:figProd}
\end{figure}
}

\def\figDecay{
\begin{figure}[htbp]
\begin{center}
\leavevmode
\hbox{
 \epsfxsize=0.40\textwidth  \epsfbox{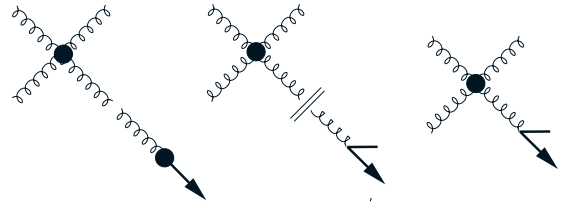}
}
\end{center}
      \caption{
a)~Generic leading-order diagram for heavy-flavor fragmentation (LO-HF),
   $\sigma(gg\to gg) \otimes D_{g\to Q}$.
b)~Subtraction diagram for heavy-flavor fragmentation (SUB-HF),
   $\sigma(gg\to gg) \otimes {}^1d_{g\to Q}$.
c)~Next-to-leading-order diagram for heavy-flavor creation (NLO-FC).
\null\hfill\null}
   \label{fig:figDecay}
\end{figure}
}
\def\figFeSub{
\begin{figure}[htbp]
\begin{center}
\leavevmode
 \hbox{
 \epsfxsize=0.22\textwidth  \epsfbox{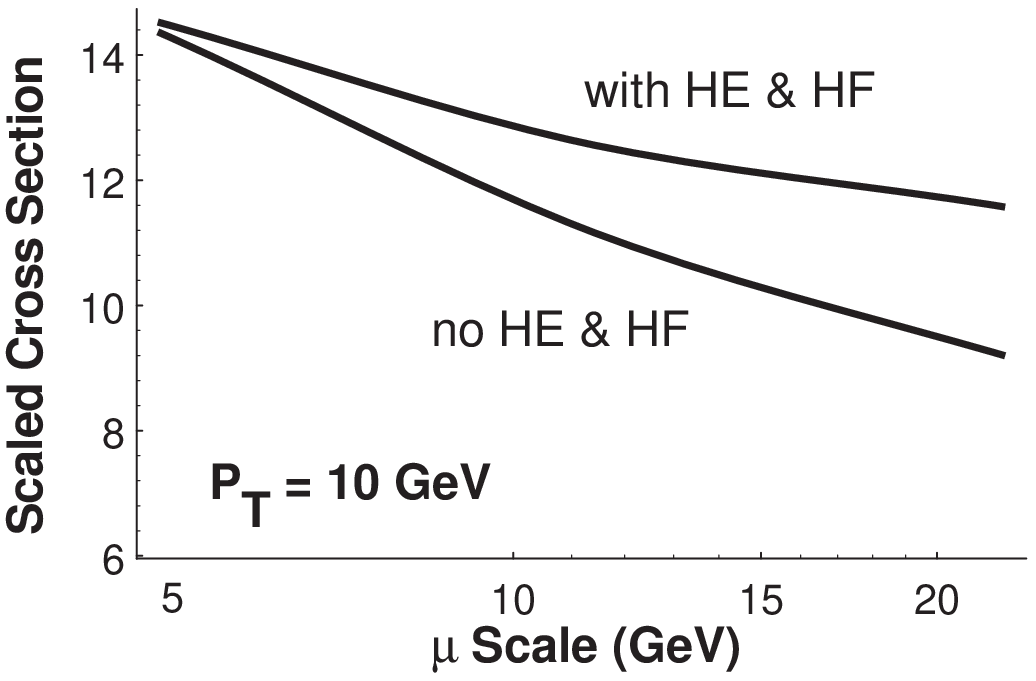}
 \hfill 
 \epsfxsize=0.22\textwidth  \epsfbox{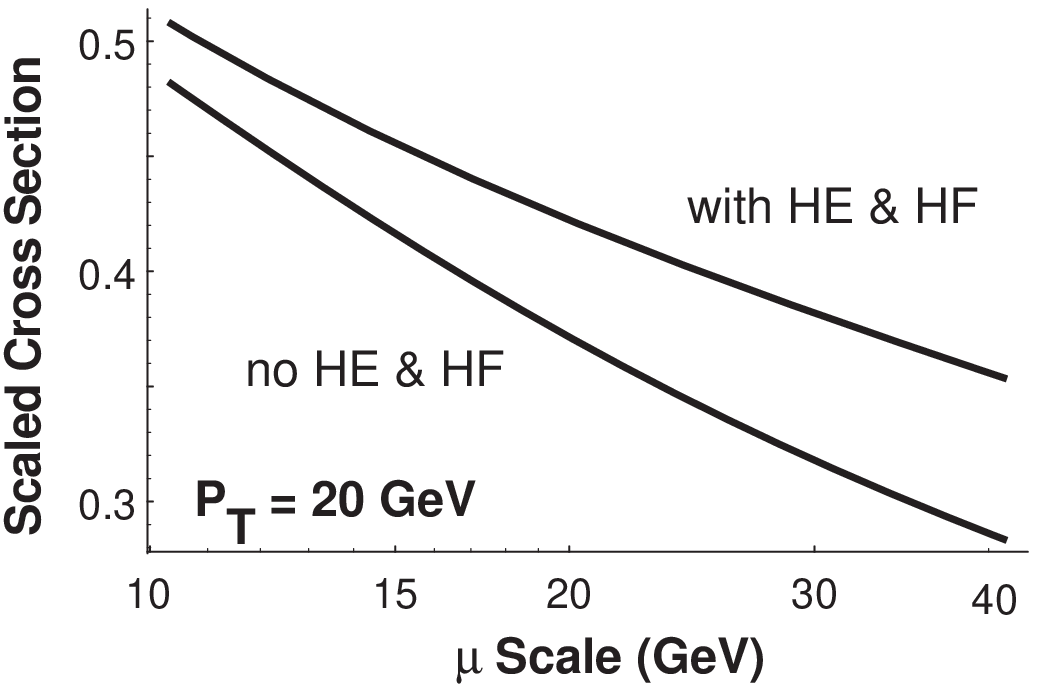}
}
\end{center}
      \caption{
The scaled differential cross section $p_T^5 \, d^2 \sigma/dp_T^2/dy$ at
$p_T=10, \, 20 \, \GeV$ and $y=0$ in  $(pb-\GeV^3)$ {\it vs.} $\mu$.
 The lower curves are the heavy quark
production cross sections {\it ignoring}
heavy-flavor excitation (HE) and heavy-flavor fragmentation (HF).
 The upper curves are the heavy quark
production cross sections {\it including}
HE  and HF. {\it Cf.}, Ref.~\protect\cite{cost}.
\null\hfill\null}
   \label{fig:figfesub}
\end{figure}
}


Improved experimental measurements of heavy quark hadroproduction has
increased the demand on the theoretical community for more precise
predictions \cite{hqdata,nde,acot,Greco,cost}. 
The first Next-to-Leading-Order (NLO)
calculations of charm and bottom hadroproduction cross sections were
performed some years ago \cite{nde}. As the accuracy of the data
increased, the theoretical predictions displayed some shortcomings:
 1) the theoretical cross-sections fell well short of the measured values,  
 and 
 2) they displayed a strong dependence on the unphysical renormalization
scale  $\mu$. 
Both these difficulties indicated that these predictions were missing
important physics. 

 \figHQdata
 \figProd 
 \figDecay 
 \figFeSub

These deficiencies can, in part, be traced to 
large contributions generated by logarithms associated with the heavy quark 
mass scale, such as\footnote{Here, $m_Q$ is the heavy quark mass, $s$ is
the energy squared, and $p_T$ is the transverse momentum.}
$\ln(s/m_Q^2)$ and $\ln(p_T^2/m_Q^2)$.  Pushing the calculation to one
more order, formidable as it is, would not necessarily improve the situation 
since these large logarithms persist to every order of perturbation theory.
Therefore, a new approach was required to include these logs.
 
In 1994, Cacciari and Greco\cite{Greco} observed that since the heavy
quark mass played a limited dynamical role in the high $p_t$ region,
one could instead use the massless NLO jet calculation convoluted with
a fragmentation into a massive heavy quark pair to compute 
more accurately the production cross section in the region $p_t \gg m_Q$. 
In particular, they find that the dependence on the renormalization scale
is significantly reduced.

A recent study\cite{cost} investigated using initial-state heavy quark
PDF's and final-state fragmentation functions to resum the large
logarithms of the quark mass.  The principle ingredient was to include
the leading-order heavy-flavor excitation (LO-HE) graph 
(Fig.~\ref{fig:figProd})
and the leading-order heavy-flavor fragmentation (LO-HF) graph
(Fig.~\ref{fig:figDecay}) in the traditional NLO heavy quark
calculation \cite{nde}. These contributions can not be added naively to
the ${\cal O}(\alpha_s^3)$ calculation as they would double-count
contributions already included in the NLO terms; therefore, a
subtraction term must be included to eliminate the region of phase
space where these two contributions overlap.  This subtraction term
plays the dual role of eliminating the large unphysical collinear logs
in the high energy region, and minimizing the renormalization scale
dependence in the threshold region.  The complete calculation
including the contribution of the heavy quark PDF's and fragmentation
functions 1) increases the theoretical prediction, thus moving it
closer to the experimental data, and 2) reduces the $\mu$-dependence
of the full calculation, thus improving the predictive power of the
theory. (Cf., Fig~\ref{fig:figfesub}.)

In summary, the wealth of data on heavy quark hadroproduction will
allow for precise tests of many different aspects of the theory,
namely radiative corrections, resummation of logs, and multi-scale 
problems.  Resummation of the large logs associated with the
mass is an essential step necessary to bring theory in agreement with
current experiments and to make predictions for the VLHC.

\subsection{W Mass Studies}

\def\figMWMT
{
\begin{figure}[bht]
 \epsfxsize=\hsize
 \centerline{\epsfbox{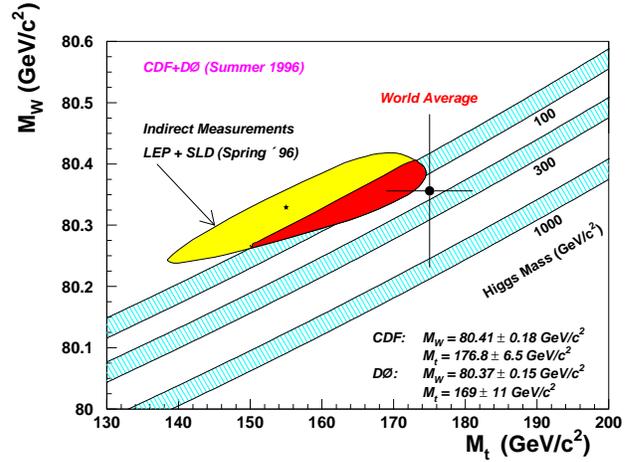}}
 \caption{
Plot of $M_W$ {\it vs.} $M_t$ with \Dzero\ and CDF preliminary measurements of
the W boson and top quark masses. Bands indicate the Standard Model
constraints for different Higgs mass values. Indirect measurements
from LEP I are also shown.
(June, 1997) 
Taken from Ref.~\protect\cite{baur_demarteau}.}
\label{fig:figMWMT}
\end{figure}
}
\figMWMT

The W boson mass is one of the fundamental parameters of the standard model;
its precision measurement can be used in conjunction with the top mass 
to extract information on the Higgs boson mass. 
 The W boson mass has already been measured precisely, and 
the current world average is:
$M_W = 80.356 \pm 0.125 \, \GeV/c^2$.

Here, we focus on issues which are unique to a VLHC facility, and
refer the reader to the literature for details regarding other 
topics \cite{demarteau_354,demarteau_353,baur_demarteau,ewreport}.
 The question addressed in the working group session was to consider 
the expected precision for $M_W$ at the VLHC in comparison to what will be 
available from competing facilities at VLHC turn-on. 
 For our estimates, we use 
 $\sqrt{s}= 100$ TeV, 
 $\Delta t = 16.7$ ns (the bunch spacing), 
 $\sigma_{tot} \simeq 120$ mb,
 and 20 interactions per crossing. 

For W events produced in a hadron collider environment there are
essentially only two observables that can be measured: {\it i)} the
lepton momentum, and {\it ii)} the transverse momentum of the recoil
system. The transverse momentum of the neutrino must be inferred from
these two observables.  The W boson mass can be extracted from either
the lepton transverse momentum distribution, or the transverse mass:
$M_T = \sqrt{2 p_T^e p_T^{\nu} (1-\cos \phi^{e\nu})}$, where
$\phi^{e\nu}$ is the angle between the electron and neutrino in the
transverse plane.

It is important to note that the following estimates necessitate a
large extrapolation from $\sqrt{s}= 1.8$ TeV to $\sqrt{s}= 100$
TeV.  For the W decays, the observed number distribution in
pseudorapidity ($\eta$) can be estimated by scaling results from the
CERN $Sp\bar{p}S$ and the Fermilab Tevatron.  The shoulder of the
pseudorapidity plateau is $\sim$ 3 for $\sqrt{s}= 630 \, \GeV$, and
$\sim$ 4 for $\sqrt{s}= 1.8$ TeV.  This yields an estimate in the
range of $\sim 5$ to 9 for a $\sqrt{s}= 100$ TeV VLHC.  Assuming
coverage out to $|\eta| \leq 4$, we obtain $\sim 1400$ charged tracks
in the detector calorimeter with which we must contend for the missing
$E_T$ calculation, ($\Emiss$).  Scaling the $\langle p_T \rangle$ up
to $\sqrt{s}= 100$ TeV we estimate $\langle p_T \rangle \simeq 865
\, MeV$ for minimum bias tracks.  Assuming $N_{ch}/N_{\gamma}=1$
yields an average $E_T$ flow of $2$ TeV in the detector.  Using
current $\Emiss_T$ resolutions of $\sim 4 - 5 \, \GeV$, we estimate
$\sigma(\Emiss_T) \simeq 25 - 30 \, \GeV$ for VLHC.

Two fundamental problems we encounter at a VLHC are multiple
interactions and pile-up.  Multiple interactions are produced in the
same crossing as the event triggered on.  The effects are
``instantaneous;" {\it i.e.}, the electronic signals are added to the
trigger signals and subjected to the same electronics.  Pile-up effects
are out-of-time signals from interactions in past and future buckets
caused by ``memory" of the electronics. Both cause a bias and affect
the resolution, but in different ways. The effect of pile-up is
strongly dependent on the electronics used in relation to the bunch
spacing.

The bottom line is the estimation of the total uncertainty on the W
mass, $\delta M_W$.  For a luminosity of 2~fb$^{-1}$, $\delta M_W$ is about 
20~MeV for both the transverse mass and lepton transverse momentum fits.
For an increased luminosity of $10fb^{-1}$, the transverse mass fit might
improve to $\delta M_W \sim 15 \, MeV$, with minimal improvement for
the determination from the lepton transverse momentum distribution.
It should be noted that these estimates have quite a few
caveats---additional study would be required before taking these
numbers as guaranteed predictions.  
In Table~\ref{tab:deltamw}, we compare these estimations
with the anticipated uncertainty from upcoming experiments.
Clearly the VLHC will not greatly improve the determination of
$M_W$.   The situation becomes more difficult when one insists that
the VLHC detectors be capable of {\em precisely} measuring  the relatively low 
energy leptons from the $M_W$ decay.

\begin{table}[htbp]
\begin{center}
\caption{
Anticipated limits on  $\delta M_W$ from present and future facilities.
(This compilation is 
taken from Ref.~\protect\cite{baur_demarteau}.)
}
\label{tab:deltamw}
\begin{tabular}{||l|c|r||} \hline \hline 
{\sc Facility }   &  $\delta M_W$ {\small ($MeV/c^2$)}  & ${\cal L}$  
\quad\quad  
\\ \hline
\hline NuTeV    & $\sim$ 100  &  --- \quad\quad  \\ \hline
HERA     & $\sim$ 60  &  1000 $pb^{-1}$   \\ \hline
LEP2	    &  $\sim$ 35-45  &  500 $pb^{-1}$   \\ \hline
Tevatron & $\sim$ 55 & 1 $fb^{-1}$   \\ \hline
Tevatron & $\sim$ 18 & 10 $fb^{-1}$   \\ \hline
LHC      & $\lsim$ 15 & 10 $fb^{-1}$   \\ \hline
VLHC     & $\sim$ 20  & 1 $fb^{-1}$   \\ \hline
VLHC     & $\sim$ 15  & 10 $fb^{-1}$   \\ \hline  \hline
\end{tabular}
\end{center}
\end{table}

\subsection{The Top Quark}

The mass of the recently discovered top quark is determined
by the CDF and
\Dzero\  collaborations from $t \bar{t}$ production at the Tevatron.
 For the details of this discovery and measurement, we refer the
reader to Refs. ~\cite{pbarp,observetop,tmass,dilepton}.

In Table~\ref{tab:top}, we display the anticipated accuracy on the top 
quark mass at the Tevatron as estimated in the TeV2000 
report ~\cite{tev2000}.
Since this report, statistical techniques have been improved such that
one would expect a precision of $\delta m_t \sim  1.5$ GeV  with $10~
fb^{-1}$,  assuming other sources of systematics are negligible. 

Moving on to the LHC, the top production cross section is $\sim 100$ times 
greater than at TeV2000, so with a luminosity of $\sim 100 fb^{-1}/year$, 
we  expect $\sim 1000$ more top events after one LHC year. 
Assuming naively that the errors scale as $1/\sqrt{N}$
(where N is the number of events), we would obtain 
$\delta m_t \sim 50$ MeV. 

The challenges of the VLHC are quite similar to the LHC regarding this
measurement.  A precision measurement of the top quark mass at this
level (or better) places stringent demands on the jet calibration.
Even with large control samples of $Z$ + jets and $\gamma$ + jets,
uncertainties due to the ambiguous nature of jet definitions will
persist.  The large number of multiple interactions at LHC and VLHC
complicates this analysis (in a manner similar to that discussed for
the W boson mass measurement).  Therefore, in order to improve upon
existing measurements, the VLHC detectors will need to be extremely
well designed and understood.

\begin{table}[htbp]
\begin{center}
\caption{
Anticipated accuracy on the top quark mass, as estimated 
by the TeV2000 report.}

\label{tab:top}
\begin{tabular}{||c|r|r|r||} \hline \hline 
Source  & 70 $pb^{-1}$ & 1 $fb^{-1}$ & 10 $fb^{-1}$   \\ \hline \hline
Statistics & 25 & 6.2 & 2  \\ \hline
Jet Scale & 11 & 2.7 & 0.9   \\ \hline
Backgrounds & 4 & 1 & 0.3   \\ \hline  \hline
Total & 27.6 & 6.9 & 2.2   \\ \hline  \hline
\end{tabular}
\end{center}
\end{table}

\subsection{Probing a nonstandard Higgs boson at a VLHC}

 We have studied the potential of a VLHC to observe a nonstandard Higgs
 boson (i.e. a spin-0 isospin-0 particle with nonstandard couplings to
 weak gauge bosons and possibly fermions) and distinguish it from the
 Standard Model  Higgs boson.
 Results are presented for different options for the energy
 ($\sqrt{s}=50, 100, 200$~TeV) and luminosity (${\mathcal L}=10^{33}-10^{35}
 cm^{-2} s^{-1}$) and compared to those obtained for the LHC in  
 ~\cite{vki}.

 Our analysis is based on
 the gold-plated channel $H\rightarrow
 ZZ \rightarrow l^+l^-l^+l^-$
 and assumes cuts on the final-state leptons, which are given by
 $|\eta^l|<3, \,p_T^l > 0.5 \times 10^{-3} \sqrt{s}$.
 We studied Higgs masses in the range from 400 to 800 GeV (600-800~GeV
 for $\sqrt{s}=200$~TeV), where the lower limit is due to the cuts and
 the upper limit is theoretically motivated. 

The two relevant parameters
 that encode the deviations from the Standard Model (SM) are $\xi$ and $y_t$,
 the $HW^+W^- (HZZ$) and $Ht\bar{t}$ couplings relative to
 the SM respectively.
 We found that a nonstandard Higgs should be detected for practically all
 values of $\xi, y_t$ and ${\mathcal L}$ in the entire mass range studied,
 a situation which is not so clear for the LHC, particularly for
 the larger masses. 

A nonstandard Higgs boson can be distinguished
 from the SM one by a comparison of its width $\Gamma_H$ and the total
 cross-section. Due to theoretical uncertainties in the latter, we chose
 to use as a criterion only the measurement of the width. Following the
 procedure of ~\cite{vki} we quantified the statistical significance
 of a deviation from the SM prediction by constructing the probability
 density function according to which the possible measurements of the
 SM width are distributed.
 Postulating that a nonstandard Higgs boson is ``distinguishable'' if its 
 width differs from the SM value by at
 least $3\sigma$, we were able to determine the precision with which
 the parameter $\xi$ can be measured at the LHC and a VLHC. This is
 summarized in Table~I for the case of $y_t=1$.
 We deduce that, for the purpose of precision
 measurements of the Higgs couplings, a lower energy VLHC with higher 
 luminosity is preferred to that of a higher energy with lower luminosity ---
 a conclusion that is due to the low-mass 
 character of the physics of interest.

Consequently, we find that for   Higgs masses in the
 range from 400 to 800 GeV,  the Higgs-Z-Z coupling
 can be measured to within a few percent at the VLHC, depending on the 
 precise mass and collider parameters.

 %
 %
 %
 %
 %
 %
 %
 %

\subsection{Supersymmetry}

Supersymmetry (SUSY) is a dominant framework for formulating physics
beyond the standard model in part due to the appealing
phenomenological and theoretical features.  SUSY is the only possible
extension of the spacetime symmetries of particle physics, SUSY easily
admits a massless spin-2 (graviton) field into the theory, and SUSY
appears to be a fundamental ingredient of superstring theory.  Given
the large number of excellent recent reviews and reports on 
SUSY \cite{lykken,reportsup,lhcsusy}, 
we will focus here on the issues directly related to the
VLHC.

One specific question which was addressed in the working group meeting was:
Is the VLHC a precision machine for standard weak-scale SUSY
with sparticle masses in the range 80 GeV to 1 TeV?
 Probably not, for the following reasons.
 \begin{itemize}
 \item
An order of magnitude increase in sparticle production rates 
will yield minimal gains, {\it except} for sparticles in the range
$\gsim$ 1 TeV. 
 \item 
Multiple interactions, degraded tracking, calibration, and b-tagging 
issues complicate reconstruction of the SUSY decay chains. 
 \end{itemize}
 \noindent
On the contrary, VLHC looks best if SUSY has some heavy surprises
such as $\gsim$ 1 TeV squarks, or $\sim$ 10 TeV SUSY messengers. 

 One example of a plausible SUSY scenario would be  heavy first and second 
generation squarks and sleptons (to suppress FCNC's) with a characteristic
mass in the range of $\sim$ 3 TeV ~\cite{lhcsusy}. 
While the gauginos and the third generation squarks and sleptons would be 
within reach of the LHC, investigation of 
 $\{ 
  \widetilde{u},
  \widetilde{d},
  \widetilde{e},
  \widetilde{\nu_e},
 \}$
and 
 $\{ 
  \widetilde{c},
  \widetilde{s},
  \widetilde{\mu},
  \widetilde{\nu_\mu}
 \}$
in the multi-TeV energy range would require a higher energy facility
such as the VLHC.  

 An estimate of the heavy squark signal over the weak-scale SUSY
background and conventional channels (such as $t\bar{t}$) indicates
that a VLHC can observe heavy quarks in the $\sim$ 3 TeV mass range;
such a heavy squark is difficult to reach at the LHC.  One might
expect on order of $10^3 - 10^4$ signal events/year. Of course,
background rejection is a serious outstanding question, and the
efficiently of b-tagging and high $p_t$ lepton detection, for example,
are crucial to suppressing the backgrounds.

\subsection{Conclusions}

While these individual topics are diverse, there are some common themes
we can identify with respect to a VLHC machine. 
  First,  a very high energy hadron collider does not appear to be the
machine of choice for precision measurements in the energy range
$\lsim$500~GeV. The competition from Tevatron, HERA, LEP, and LHC are
formidable in this region. To obtain comparable precision, the
VLHC is handicapped by numerous factors including 
multiple interactions, 
large multiplicity, and large $\Emiss$. 

In contrast, the strong suit of the VLHC is clearly its kinematic 
reach. Should there be unexpected sparticles in the $\gsim$~TeV 
range, the VLHC would prove useful in exploring this range. 
 Of course our intuition as to what might exist in the $\sim$10~TeV
regime is not as refined as the $\lsim$1~TeV regime which will
be explored in the near-future; however what
we discover in this energy range can provide important clues as
to where we should search with a VLHC.

\clearpage

\section{Multiple Interactions Working Group}
\begin{center}
{G. Snow\\
{\it University of Nebraska-Lincoln}\\
\ F. Paige\\
{\it Brookhaven National Laboratory}}\\
\end{center} 

 The Multiple Interactions working group was charged with investigating
 issues related to the large number of
 interactions per crossing envisioned for the VLHC at design
 luminosity. Presentations and discussions focused on the following
 topics: corrections to the calorimetric measurements
 of single-particle and jet energies in the presence of many
 interactions, multiple interaction corrections to luminosity
 measurements at the Tevatron, and the measurement of the luminosity in
 the VLHC environment. As the Tevatron luminosity increased into the
 range 10$^{31}$ cm$^{-2}$ s$^{-1}$ during Run I, the CDF and D\O\
 experiments learned to cope with increasing, yet small number (1 -- 5) of 
 interactions per crossing. The LHC at CERN will provide a more 
 difficult training ground since there will be in average 17 interactions per 
 crossing at the design luminosity of 10$^{34}$ cm$^{-2}$ s$^{-1}$.

 \begin{figure}[h]
 \leavevmode
 \resizebox{8cm}{!}{%
 \includegraphics{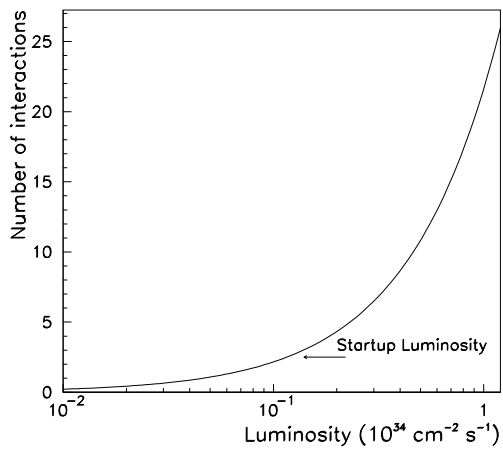}}
   \caption[]{Average number of interactions per crossing as a function of
 luminosity at the VLHC. The horizontal axis is a log scale labeled in
 units of 10$^{34}$ cm$^{-2}$ s$^{-1}$.} 
   \label{fig:intvslum}
 \end{figure} 

 The problem with multiple interactions at the VLHC will be worse, yet
 comparable to the situation at the LHC. The design luminosities are
 identical (10$^{34}$ cm$^{-2}$ s$^{-1}$) and the luminous region for a
 given bunch crossing will be a few cm longitudinally for both
 machines. While the time between bunch crossings is 25 nsec at the 
 LHC and the bunches are separated by 7.5 m, these numbers are 17 nsec 
 and 5 m, respectively, for the VLHC. Fig.~\ref{fig:intvslum} 
 shows the number of
 interactions per bunch crossing expected at the VLHC as a function of
 luminosity, assuming an inelastic proton-proton cross section of 130
 mbarn. At design luminosity, each beam crossing will yield about 
 22 interactions. 

 Both the LHC and the VLHC will likely come online with instantaneous
 luminosities at least a factor of 10 lower than the design luminosity.
 Fig.~\ref{fig:intvslum} shows that at start-up luminosity, there will only be a few
 interactions per crossing, so the multiple interaction problem will be
 similar to that faced at the Tevatron. As discussed below VLHC will 
 benefit from low-luminosity running at start-up,
 both for physics and detector calibration reasons. 

 The much higher center-of-mass energy of the VLHC, however, will make
 the underlying event problem more difficult than at the LHC, since the
 particle multiplicity and average minimum-bias E$_{T}$ will be higher. 
 Still an average E$_{T}$ density of 10's of GeV per unit $\eta - \phi$ 
 at the VLHC design luminosity is manageable if one is searching for high mass
 particles and jets at $\sqrt{s}$ = 100 TeV. 

 A precise knowledge of the proton-proton luminosity at a VLHC
 interaction region is an essential ingredient in the measurement of
 absolute cross sections in a VLHC experiment. Monitoring the instantaneous
 luminosity is also important for making corrections to the data for
 detector effects related to the number of interactions per beam
 crossing. 

 The "counting zeros" technique is used by the D\O\ and CDF experiments
 at the Fermilab Tevatron collider and leads to an uncertainty of order
 5\%. A modified version of this technique is expected to yield similar
 precision at the VLHC even in the presence of large numbers of
 interactions per bunch crossing. 

 The counting zeros technique works as follows \cite{ref:d0note2860}. 
 Two sets of luminosity
 monitors, symmetrically located on each side of the interaction region, 
 count the fraction of times a given bunch crossing results in no detected
 particles on either side. The luminosity is inferred from the rate of
 such zeros. 

 The probability of having an empty crossing where a
 forward/backward coincidence is not recorded is given by:
 $$ P(0) = e^{-\overline{n}_{1}}\,(2\,e^{-\overline{n}_{2}/2} -
 e^{-\overline{n}_{2}}) $$
 where 
 $\overline{n}_{1}$ is the average number of forward/backward 
 coincidences and 
 $\overline{n}_{2}$ is average number of one-side hits (but not both). 
 $\overline{n}_{1}$ and $\overline{n}_{2}$ are related to
 the instantaneous luminosity {\it L} via:
 $$ \overline{n}_{1} = (\epsilon^{sd}_{1}\sigma^{sd} +
 \epsilon^{dd}_{1}\sigma^{dd} + \epsilon^{hc}_{1}\sigma^{hc})\,\tau\,
 {\it L}
 $$
 and 
 $$ \overline{n}_{2} = (\epsilon^{sd}_{2}\sigma^{sd} +
 \epsilon^{dd}_{2}\sigma^{dd} + \epsilon^{hc}_{2}\sigma^{hc})\,\tau\,
 {\it L}
 $$
 Here, 
 $\sigma^{sd}, \sigma^{dd}, \sigma^{hc}$ are the cross sections for
 single-diffractive, double-diffractive, and hard-core scattering,
 $\epsilon^{sd}_{1}, \epsilon^{dd}_{1}, \epsilon^{hc}_{1}$ are the 
 acceptances for forward/backward coincidences for these processes, 
 $\epsilon^{sd}_{2}, \epsilon^{dd}_{2}, \epsilon^{hc}_{2}$ are the 
 corresponding acceptances for one-side hits, and 
 $\tau$ is the bunch crossing time.

 \begin{figure}[h]
 \leavevmode
 \resizebox{8cm}{!}{%
 \includegraphics{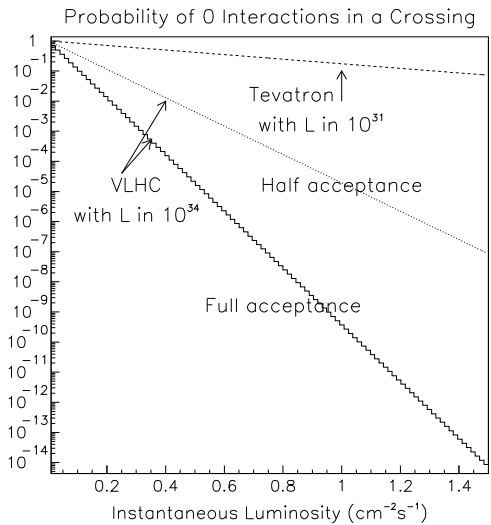}}
   \caption[]{The probability of detecting zero interactions in a crossing 
 vs. instantaneous luminosity for the Tevatron (full acceptance Level-0 
 array) and the VLHC (full and half acceptance arrays).}
   \label{fig:zerovslum}
 \end{figure}

 In order to track the luminosity as it decreases through the lifetime
 of an accelerator store (typically from a few hours to a day), for
 example, one would like to monitor the luminosity with a statistical
 uncertainty of about 1\% every few seconds or minutes. This calls for
 counting of order 10$^{4}$ zeros in this period. These rates are
 achieved in the D\O\ experiment with the fine-grained Level-0 array
 of scintillation counters \cite{ref:d0nim} which subtend the
 high-$\eta$ region on both sides of the interaction region. The
 Level-0 counters are nearly 100\% efficient for detecting a
 forward/backward coincidence from a hard-core scattering event, the
 dominant process among those listed above, since the two beam jets
 almost always send particles into the two arrays.

 \begin{figure}[h]
 \leavevmode
 \resizebox{8cm}{!}{%
 \includegraphics{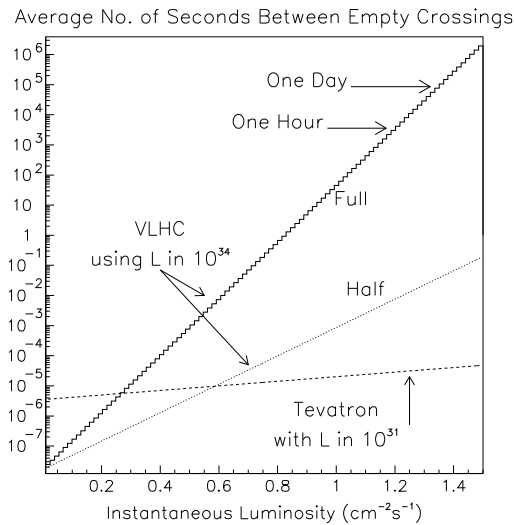}}
   \caption[]{The average number of seconds between detected empty
 crossings vs. instantaneous luminosity for the Tevatron (full
 acceptance Level-0 array) and the VLHC (full and half acceptance
 arrays).} 
   \label{fig:secvslum}
 \end{figure} 

 Fig.~\ref{fig:zerovslum} shows the probability of detecting zero 
 interactions per
 crossing as a function of luminosity for the D\O\ configuration and
 for a similar high-acceptance, high-efficiency "Level-0" array
 located in a VLHC experiment. Fig.~\ref{fig:secvslum} shows the same 
 information in
 an alternate way -- the average number of seconds between empty
 crossings vs. luminosity. One sees that a full-acceptance array at the 
 VLHC results in having to wait several minutes between empty 
 crossings at the design luminosity. The rates of detected zeros can 
 be effectively increased, however, by decreasing the $\epsilon_{1}$ and 
 $\epsilon_{2}$ terms in the above relations. This can be achieved by 
 using an array of luminosity counters which have a smaller geometric 
 acceptance or are less efficient for detecting minimum-ionizing 
 particles, accomplished by raising discriminator thresholds. 

 Fig.~\ref{fig:intvslum} and~\ref{fig:zerovslum} also show that the
 VLHC situation for a half-acceptance array starts to approach that of
 the Tevatron. Extrapolating further, the figures indicate that
 acceptance terms of order 10\% those used at the Tevatron will result
 in zero-counting rates which give negligible statistical uncertainty
 in the luminosity measurement at the VLHC design luminosity. Fast
 timing for such counters will be necessary to distinguish between the
 bunch crossings separated by 17 nsec. Tevatron Level-0 counters, with
 200 psec time resolution, have already demonstrated the ability to
 distinguish particles in neighboring buckets which are separated by
 about 15 nsec.

 The calibration of the luminosity counters at the VLHC will require
 running the VLHC at low luminosity, where there is an average of one
 interaction per crossing. This will likely be the default scenario
 during the start-up of the machine. Many physics measurements require
 low-luminosity running as well: studies of elastic scattering and
 diffraction dissociation, studies of rapidity gaps between jets, etc.
 Another important step in calibrating the luminosity counters will be
 to run the VLHC at a lower center-of-mass energy where the total
 proton-proton cross section and its components (hard-core, elastic,
 single-diffractive, and double-diffractive) have been accurately measured. 
 The LHC
 center-of-mass energy of 14 TeV would be the obvious low-energy
 target. Hence, the VLHC machine designers should incorporate into
 their planning the possibility of running the machine stably at this
 energy. Running the VLHC at the LHC center-of-mass energy will be
 useful for cross checking other ingredients in cross-section
 calculations besides luminosity, as well as studying the
 $\sqrt{s}$-dependence of many physics processes.

 \clearpage

\section{Tracking Working Group}

\begin{center}
{F. Borcherding\\
{\it Fermi National Accelerator Laboratory}\\
\ Tao Han\\
{\it University of California--Davis}}\\
\end{center} 

\subsection{Introduction}

Good tracking has been and will continue to be a key ingredient for high energy
physics experiments.  Good tracking will require an inner tracker which can
achieve precise measurements of the vertex positions for both the initial
and displaced vertices.  This good tracking will also require an outer
tracker which can supply a precision position at large lever arm for
momentum measurements and which can be used as a seed for track finding
into the inner tracker.  The tracker must be resistant to radiation
damage particularly in its inner layers which are closest to the beam.
\par
The machine parameters specify 100 TeV center of mass energy and a peak 
luminosity of $10^{34} cm^{-2}sec^{-1}$. With a bunch spacing of 17ns there 
will be about 20
interactions per crossing.  Extrapolations of the minimum bias cross
section are uncertain but indicate that each collision will generate 
about $10^{2}$ particles for central ($\eta$ $<$ 1.5)
and of the same order for each of forward and backward 
(1.5 $<$ $\eta$ $<$ 3) regions.  This produces
some 4x$10^{3}$ charged tracks on average per crossing. 
For tracking to work well the occupancy needs to be kept to about 1\% which
leads to a requirement of  4x$10^{5}$ channels per tracking layer.
\par
It is important to have a precision determination of the momenta of very
high energy charged leptons (as discussed by T. Han \cite{t_1}).  For one 
meter of tracking in a four Tesla field, using fifteen planes, each with 
50 $\mu m$ resolution, the $\frac{\sigma_{p}}{p}$
resolution for a 10 TeV particle is 25\%.  For two meters of tracking this
drops to 10\%.  Getting to 2.5\% in one meter requires 5 $\mu m$ resolution. 
Remember that 50 $\mu m$ resolution is the 'goal' for the LHC detectors.  Also
needed is effective second vertex detection.  For a two plane system the
impact parameter resolution is a function of the ratio of the inner to
outer radius.  Therefore the inner radii must be small, close to the beam,
or the outer radii becomes very large.

\subsection{Inner Tracking with Pixels}

For the high radiation levels and instantaneous rates of the VLHC the best
choice for inner tracking appears to be pixels 
(as discussed by S. Kwan \cite{t_2}).  Even at $\sqrt{s}$ = 100 TeV but
with a luminosity of about that at the LHC the requirements placed on the
pixels by rate considerations are about the same.  There pixel detectors of
the order of $10^{8}$ channels are planned with r-$\phi$ 
hit resolutions of about
15 $\mu m$.  But at the VLHC the momentum of the high momentum tracks will be
almost a factor of ten higher.  One way to preserve momentum resolution is to
improve tracker resolution by a factor of ten.  Some work has been done which suggest
that this factor of ten is possible, but reading out the many more smaller
pixels could remain a problem \cite{t_3}.

\subsection{Outer Tracking with Gas Chambers}

A promising approach to solving these difficult problems appears to  be
the combination of the tried and true proportional mode gas avalanche
counter and new techniques of surface treatments and photo lithography ( as discussed by P. Rubinov \cite{t_4}).
This is already a very active field which has produced ideas such as the
Micro Strip Gas Chambers (MSGCs), Micro Gap Chambers (MGCs), Gas Electron
Multiplier (GEM) and more.  Already one of the LHC detectors (CMS) has
committed to developing the MSGCs for tracking as the most promising path
of achieving the required performance.  For the VLHC, only incremental
improvements beyond the LHC application would be required.
Gas detectors created with micro technology offer the following parameters:
\begin{itemize}
\item Spatial resolution - 40 $\mu m$ is achievable without difficulty.
\item Timing resolution - The very short charge collection times allow a timing
	 resolution of approximately 10ns with current technologies.
\item Low mass - Intrinsically, a 3mm gas gap at 1 atm is sufficient for the 
         detection of minimum ionizing particles
	 with full efficiency, but in practice all the material is in the
	 support.
\item High rate capability - Up to $10^{6}$ particles/$mm^{2}$/sec is currently 
	 achievable (r $>$ 190mm at VLHC).
\item Segmentation - Strips of as long as 30cm can be used, or they can be as
	 small as 200 $\mu m$ for MSGCs or even less for MGCs.
\item Radiation hardness - Able to withstand several megarads of ionizing 
	 radiation.  
With current technologies, MSGCs can be made to work for up 
	 to 10 years in an LHC environment.
\end{itemize}
\par
Gas detectors is a very active field that is in the process of rapid evolution. 
Some promising ideas, such as the GEM are only now beginning to be
explored.  Currently the following parameters can be achieved
simultaneously using MSGCs:
\begin{itemize}
\item 40 $\mu m$ spatial resolution
\item $10^{6}$ particles/$mm^{2}$/sec rate capability
\item up to 100mC/cm of collected charge without aging with good control
		 of the gas purity.
\item 11ns RMS time resolution.
\end{itemize}
\par
Further developments such as combining the MSGC with the Gas Electron
Multiplier are expected to extend the performance by allowing much higher
gains and significantly improved reliability as well as extending the rate
capability by another order of magnitude.

\subsection{Scintillating Fibers}

Scintillating fibers are a viable technology at high luminosity
(as discussed by F.Borcherding \cite{t_5}).  A fiber
tracker could start at an inner radius of 1.6m if that layer is segmented
into 10 sections in z and the fiber diameter is as small as 0.5mm.  A fiber
tracker for the VLHC then could have eight layers each with 2x$10^{5}$ 
channels located at radii of 1.5 to 3.0m.
\par
The Upgrade D\O\ detector at FNAL will use VLPCs to convert photons seen in
its fiber tracker and has a total channel count of just under $10^{5}$ in
eight layers \cite{t_6}.  It is not unreasonable to expect to build a detector eight
to ten times as large for about the same cost in over a decade from now.  A
modest extrapolation of the VLPC technology indicates that the channel
count could be increased 4-fold with no increase in the number of VLPC
chips.  There is some evidence that 0.5mm diameter fibers will produce
enough photons to work with the present VLPCs \cite{t_7}.  A major factor would be in
the cost and room needed for the 16 fold increase in clear wave guide
fibers.  This would be greatly reduced if the VLPCs which must operate at
about 6.5 degrees K could be moved closer to the fiber tracker.
The electronics could be greatly streamlined over the D\O\ design by
requiring only one bit of information for each channel.  The present front
end pick off chip for D\O\ does this for 16 channels.  This chip could be
evolved up to 64 or 128 channels and a pipelined output stage added.

\subsection{3D Pixels}

Today's silicon strips and tomorrow's pixels are 2-D technology with the
electrodes etched on the surface of a silicon wafer (as discussed by 
S. Parker \cite{t_8}).  In 3-D technology the
electrodes would extend through the wafer thickness.  Here the n- and
p-strips instead of laying on the surface are columns extending through the
300 $\mu m$ wafer thickness.  The depletion voltage for 3-D pixels is very much
smaller than that for 2-D pixels.  The signal amplitude is also very much
greater and arrives within 1ns.
The chip industry in the past has been focused on surface features.  In the
future, however, it will probably move into 3-D structures in its quest for
denser and denser circuits.  In such an industry 3-D pixel manufacture
could become economically viable.

\clearpage 
\section{Calorimetry Working Group}
\begin{center}
{D. Khazins\\
{\it Duke University}\\
\ J. Lykken\\
{\it Fermi National Accelerator Laboratory}}\\
\end{center}

\subsection{Introduction}

The calorimetry looks as the most feasible part of the
detector, even with the VLHC luminosity of $10^{34} cm^{-2}s^{-1}$. 
The
radiation doses in the calorimeters and their occupancies increase very
modestly with the collider energy. Based on the phenomenology developed by D.E.
Groom \cite{c_1}, 
the calorimetry radiation doses at the collider energy of 100 TeV are only 2 times higher of those at the LHC with the  same luminosity.
Similar conclusions are followed from N. Mohov's calculations \cite{c_2}. 
There are many calorimetry techniques which potentially fit the VLHC
conditions, as it was demonstrated at this Workshop. 

\subsection{Scintillator calorimeters}

This well established technique can be used for the barrel part of the
detector  in spite of its limited
radiation resistance (about 4 Mrad, according to estimations of A. Pla-Dalmau
\cite{c_3}). However, chances to use scintillator for the 
forward/backward parts of
the detector (pseudorapidity $\eta > 2$) do not look realistic even at the
luminosity of $10^{34} cm^{-2}s^{-1}$. 
\par
Performance of the CMS scintillator hadronic calorimeter in combination with
the $PbWO_{4}$ crystal electromagnetic (EM) calorimeter 
was discussed by J. Freeman
\cite{c_4}. The problem with this combination is a non-uniform response to the
hadronic and EM parts of the shower (e/h $\ne$ 1) 
which causes non-linear 
amplitude versus energy dependence for
hadrons and a degradation of the hadronic energy
resolution (increase of the constant term). However, this degradation  is
relatively small, and author concludes that it is less important than
non-gaussian tails in  the calorimeter response function due to cracks and dead
areas in real calorimeters.

\subsection{$PbWO_{4}$ crystal calorimeters}

The CMS crystal EM calorimeter was reported by R. Rusack \cite{c_5}. The lead
tungstate crystals ($PbWO_{4}$) 
allow to construct a compact EM calorimeter (the
radiation length is 0.89 cm, Moliere radius is 2 cm) with excellent energy
resolution:
\begin{center}
$\frac{\sigma_{E}}{E} = \frac{2\%}{\sqrt{E}} \oplus 0.5\% \oplus 
\frac{0.15}{E}$
(E in GeV).
\end{center}
The long term radiation resistance of the crystal is also good, 10 Mrad has
been demonstrated. Nevertheless, the ability of the $PbWO_{4}$
 calorimeter to work
in high radiation fields  still is not clear. First, there is so called `short
term' radiation damage which may vary the  crystal light output in some
unpredictable way, thus deteriorating the calorimeter resolution. Another point
of concern is the radiation resistance of the silicone avalanche photodiode 
(APD), which was accepted by CMS as a photodetector for the crystals.
\par
Both, the $PbWO_{4}$ crystals and especially the APD, are very sensitive to the
temperature variations. However, with the proper temperature monitoring,
corrections could be made and  the calorimeter energy resolution should not
suffer.
\par
Apparently, we will have more data on this promising technique in the near
future due to intensive studies for CMS.

\subsection{Quartz fiber calorimeters} 

This new type of calorimetry was presented by O. Ganel \cite{c_6}.
 The
calorimeter is a kind of `spaghetti' calorimeter with fibers made of quartz
(amorphous silica) instead of scintillator plastic. Pure quartz is very
radiation hard, 30 Grad is achievable. Quartz fibers detect
Cherenkov light which yields low-intensive but extremely  fast signal, an
output signal of several nanosecond width has been observed. 
\par
The calorimeter EM energy resolution is determined by the photo-electron
statistics (yield is ~ 0.8 ph.e./GeV):
\begin{center}
$\frac{\sigma_{E}}{E} = \frac{(100 - 140)\%}{\sqrt{E}}$.
\end{center}
Since the quartz fibers pick up only Cherenkov light from fast electrons the
calorimeter is practically non-sensitive to the hadronic energy. Nevertheless,
it can be used for the hadron energy measurements at very high energies, due to
high EM component in the  hadronic shower which logarithmically increases with
energy. Clearly, the hadronic energy resolution is not very good, basically, it
is 2 - 3 times worse than that for the conventional  (scintillator, liquid
argon) calorimeters. 
\par
Another feature of the quartz fiber calorimeter is narrow hadronic shower
(64 mm  diameter), because it detects the EM core of the hadronic
shower only. This feature may  be used for better jet-jet resolution at very 
high
rapidity regions where radius of jet cone is less than the width of the hadronic
shower.

\subsection{Diamond calorimeters}

Diamond detectors (presented by R. Stone-Rutgers \cite{c_7}) 
allow to construct very
compact, radiation hard ($\rangle$ 100 Mrad), 
robust, and very fast ($\sim$ 1ns readout)
sampling calorimeters. One such calorimeter has been constructed and tested
with a reasonable EM  energy resolution:
\begin{center}
$\frac{\sigma_{E}}{E} \sim  \frac{20\%}{\sqrt{E}} \oplus 1.5\% \oplus 
\frac{8\%}{E}$.
\end{center}

The main problem with this technique is the diamond price. Presently, the cost
of the  diamond calorimeter is one or two order of magnitude higher than other
types of calorimeters. However, the technology development may change the
situation.

\subsection{High pressure tube gas calorimeters}

Although the first calorimeter with gas ionization readout has been tested
back in 1979 \cite{c_8} this 
technique has never been used in physics experiments by
one reason: small signal. At low energies electronic noise dominates
calorimeter energy resolution. However, with the energy increase and with the
electronics improvement this technique presents  very attractive option for the
future colliders \cite{c_9}. 
\par
Gas ionization calorimeters are very radiation hard ($\rangle$ 1 Grad) 
and fast (20 ns
total width output signal has been demonstrated \cite{c_10}). 
Due to high ion mobility
in gases these calorimeters can work in high intensity radiation fields (100
rad/s) without signal degradation. Due to the lack of the gain the
calorimeters are linear and stable. Finally, the
gas ionization calorimeters are inexpensive, since their basic
design is carbon steel tubes filled with argon based gas mixtures.
\par
The tube design of the high-pressure (100 atm) gas-ionization calorimeters was
presented by D. Khazins \cite{c_11}. 
A tested hadronic calorimeter made of 0.5 inch
diameter tubes had the energy resolution: 
\begin{center}
$\frac{\sigma_{E}}{E} = \frac{70\%}{\sqrt{E}} \oplus 7.4\%$.
\end{center}

Using a weighting procedure for the e/h compensation, authors managed to reduce
the constant term to $\langle$3\%. (It is not clear, however, how this or similar
procedure could be beneficial in the case of jets.) The electronic noise of the
tested calorimeter was equivalent to 4 GeV (r.m.s) per hadronic shower. Authors
believe it could be reduced to 1 GeV or less.

An EM calorimeter made of `wiggling' tubes had the energy resolution:
\begin{center}
$\frac{\sigma_{E}}{E} \sim \frac{32\%}{\sqrt{E}} \oplus 3\%$.
\end{center}
It is somewhat worse than the expected value of $\frac{20\%}{\sqrt{E}} \oplus$
 1\%, which implies
that the wiggling calorimeter needs more R\&D work. The EM calorimeter 
electronic noise was 0.3 GeV. 

\subsection{Moderate pressure gas calorimeters with planar electrode geometry}

Another approach to the gas ionization calorimeters is being developed by a
Serpukhov  group (S. Denisov et al. \cite{c_12}). 
They employ the standard sandwich
geometry. A hadronic calorimeter filled with 90\%Ar + 10\%CF$_{4}$ 
gas mixture at 40
atm pressure has been tested. They are now looking for heavy carbon-fluorine
gases to reduce the pressure to several atmospheres.
\par
In the process of the calorimeter investigation the group discovered that they
can control the calorimeter e/h ratio by adjusting the width and delay of the
ADC gate signal. As the result they obtained a very low constant term in the
hadronic energy resolution: 2.5\% with the absorber made of steel and 0.1\% with
the lead absorber. This discovery opens possibilities for a really good
hadronic and jet energy 
resolution at VLHC, because at those energies the constant term
will dominate energy resolution.

\subsection{Liquid argon calorimeters }

The liquid argon calorimeters have not been presented at the
Workshop.  However, this well established and solid technique, undoubtedly,
would be one of the main  options for the VLHC detector. It is intrinsically
radiation hard, linear, and stable. Energy  resolution is very good for both
hadrons and EM particles. The H1 group \cite{c_13}
 obtained the  hadronic resolution:
\begin{center} 
$\frac{\sigma_{E}}{E} = \frac{51\%}{\sqrt{E}} \oplus 1.6\%$,
\end{center}

using a weighting procedure for the e/h compensation. (Again, the low constant
term may be not applicable for jets because of difference in energy
distribution in jets and hadronic showers.) 

The ATLAS group \cite{c_14} has tested 
an accordion EM calorimeter with the resolution:
\begin{center}
$\frac{\sigma_{E}}{E} = \frac{7.7\%}{\sqrt{E}}$
\end{center}

with a negligible constant term.
\par
The drawback of the liquid argon technique is the low mobility of both
electrons and ions. The big electron collection time, which is presently about
0.5 $\mu$sec, creates serious pileup problems for calorimetry. 
 The positive ions build up a volume charge in
the liquid argon gap which considerably distorts electric field in the gap at
the dose rate about 1 rad/s. However, both these limitations strongly depend
on the calorimeter design parameters (the gap size and voltage) and could  be
improved.

\subsection{Calorimeter in situ calibration}

At the last talk of the calorimetry group R. Vidal \cite{c_15} 
considered several processes for the calorimeter in situ calibration. Decays 
J/$\psi \rightarrow \mu^{+}
\mu^{-}$ and Z $\rightarrow e^{+} e^{-}$ can be  
used for the EM calorimeter calibration. The hadronic
calorimeters may be calibrated with (Z + jets) events and with b-tagged
W-bosons (decaying into two jets) from tt-bar events. The challenge is the in
situ calibration at energies exceeding the weak boson masses.

\clearpage
 
\section{Muon Working Group}
\begin{center}
{T. LeCompte\\
{\it Argonne National Laboratory}\\
\ M. Berger\\
{\it Indiana University}}\\
\end{center} 

\subsection{Introduction}

   Lepton identification is at the core of hadron collider
physics.  Leptons indicate the presence of an electroweak
boson, either real or virtual.  Electrons and muons (and to
a lesser extent taus) can be identified at the trigger 
level, allowing these interesting events to be selected
against the enormous background of QCD events.

   Muons are simple to identify.  Their long lifetime and high
penetrating power virtually independent of energy makes them very
distinctive: any charged particle that penetrates several meters of
material is a muon.  Unlike electrons, which can have their momenta
measured by calorimetry, muons have to have their momentum measured in
tracking, which becomes increasingly difficult at high $p_T$.  The
critical issue is not whether or not we can identify muons at the VLHC
- we can.  It's whether or not we can accurately measure their
momentum.

\subsection{Theoretical Issues: Which Muons Are We Looking For?}

   There are several processes of interest that can generate
muons as signatures:

$\bullet$ Compositeness or a new contact interaction: here one
signature is an increase in the Drell-Yan cross section at high
$m(\mu^+\mu^-)$, and $p_T(\mu)$ can be several TeV.

$\bullet$ New gauge bosons, such as $Z^\prime \rightarrow \mu^+\mu^-$.
Again, $p_T(\mu)$ can be several TeV. Additionally, measuring the
forward-backward asymmetry $A_{FB}$ of this new $Z^\prime$ provides
information on its couplings.  This technique requires large $\eta$
coverage for muons, and good resolution is needed since the asymmetry
varies with $m(\mu^+\mu^-)$.

$\bullet$ Heavy squarks and gluinos.  Weak scale supersymmetry (SUSY)
will presumably be discovered at the LHC if not before. However some
of the heavier states might be too heavy to be seen at the LHC or be
produced in insufficient numbers to allow a detailed study of the
complicated cascade decay.  These particles can be pair produced at
the VLHC, and many of these have muons as daughters, and these muons
are expected to have $p_T(\mu)$ in the 100's of GeV range.

$\bullet$ Strongly Interacting Electroweak Symmetry Breaking: A hadron
machine with the energy envisioned here is a ``gauge boson collider,''
since the production mechanisms involving electroweak gauge bosons
from the initial state partons becomes increasingly important.  Any
evidence uncovered at the LHC for a strongly interacting sector that
breaks the electroweak symmetry would motivate a higher energy machine
for study of the new interactions. Good charge determination for very
energetic muons would allow the identification of various isospin
channels in strong vector boson scattering. This is especially the
case in the mode $W^+W^+\to W^+W^+$ where the like-sign lepton signal
must be separated from the unlike-sign Standard Model background.

$\bullet$ Multi-$W$ production:  While these $W$'s 
tend to be at high $p_T$ (and therefore generate high $p_T$
muons), we would also like to measure the cross section
for single $W$, $WW$, etc. production, and this means detection in some cases
(when one of the $W$'s is not highly boosted with respect to the lab frame) 
down to a few 10's of GeV.

$\bullet$ Other new particles: Other scenarios include the possibility
of pair-producing leptoquarks which decay into a lepton and a quark
jet or pair-producing vector-like quarks which have leptonic decays.
New heavy particles exist in the messenger sector of gauge-mediated
SUSY breaking models. If light enough some of these particles could be
produced at the VLHC.  One possible signal involving muons in the
final state would be the production of a pair of (charged) messenger
scalars which decay into a $W$ and its (absolutely stable or
relatively long-lived) neutral electroweak doublet partner, i.e. $\phi
^+\to \phi^0W^+$.  More generally the presence of new particles could
enhance the number of muons observed at the VLHC.

    This brings up the question of the purpose of a VLHC experiment:
does it emphasize doing 10 TeV scale physics, or does it emphasize
doing high statistics 1 TeV scale physics?  The answer to this
question depends on what the LHC does and does not discover, but it
does have implications for detector design.  For very massive objects,
the acceptance is proportional to solid angle coverage, but for less
massive objects, it is proportional to rapidity.  A detector optimized
for 10 TeV scale physics will invest more resources into the best
momentum resolution in the central region, whereas a detector
optimized for 1 TeV physics will opt to cover a larger region, with
less emphasis on resolution.  This increases the yield, but also makes
$A_{FB}$ measurements possible.

    The best momentum measurement possible is desirable.  The better
the momentum resolution, the narrower peaks become, and the smaller
the signal that can be identified over the background.  Additionally,
should (e.g.) a new $Z^\prime$ boson be discovered, measuring the
width $\Gamma(Z^\prime)$ would be of intense interest.  For intrinsic
widths smaller than the detector resolution, this becomes extremely
difficult.

\subsection{Experimental Issues: How Do We Find Them?}   

    The dynamic range of the VLHC muon system is unprecedented,
ranging from a few 10's of GeV to a few 10's TeV.  Even if one were to
stipulate that muons from low $p_T$ $W$ bosons were uninteresting, it
is impractical to build a detector with a thick enough muon absorber
to be blind to these muons: approximately a meter of iron is required
for a muon to lose 1 GeV via $dE/dx$.

    Three strategies are commonly employed in measuring the momentum
of muons: measure the momentum in a central tracker, measure the
momentum in instrumented magnetized shielding, and measure the
momentum in a special muon spectrometer outside the shielding.

    It is unlikely that an independent muon tracker could measure the
momentum better than a central tracker.  An independent muon outer
tracker covers a substantially larger volume, which increases the
channel count for a given number of measurements of a given
resolution, and also reduces the practical magnetic field allowed
because of stored energy considerations. The tracking group believes
that an inner tracker using a large bore high field magnet and
existing tracking technologies can reach momentum resolution of 10\%
or better for a 10 TeV track, which corresponds to a $3\sigma$ charge
measurement for a 30 TeV muon.  Although improved momentum resolution
is always better, this is believed to be adequate to probe a broad
range of physics.

    It is expected that the dominant source of apparent high $p_T$
muons will be low $p_T$ muons that somehow get mismeasured to appear
to be much straighter than they really are.  While this is unlikely to
happen to any particular muon, there are so many more low $p_T$ muons
than high $p_T$ muons that non-Gaussian tails on the resolution could
pose substantial problems.  This is especially true at the trigger
level, where only a subset of the detector information is available.

    Clearly a second (and possibly even a third) momentum measurement
to confirm the central tracker is desirable.  It is certainly possible
to build a tracking system inside the absorber (as D\O\ has done) or
beyond the absorber (as ATLAS has done), at some cost.  Additionally,
properties of very energetic muons can be exploited.

    A 2 TeV muon has the same velocity as a 10 GeV electron.  Muons,
therefore, begin to show features normally associated with electrons
in their passage through matter.  For example, a 1 GeV muon deposits
(an average of) 3.5 GeV in 3m of iron.  A 1 TeV muon deposits 9 GeV.
If a VLHC detector built a 1 foot deep ``muon calorimeter'' at the
extreme outer radius of the muon detector, this calorimeter would only
need $30\%/\sqrt{E}$ resolution to provide $3\sigma$ separation.  The
main difficulty with this technique is shower fluctuations.  Thicker
calorimeters are less sensitive to fluctuations, although they are
more expensive: to reduce the fluctuations by a factor of two requires
a calorimeter four times thicker.

   A second property that can be exploited is that muons of this
velocity exhibit transition radiation, with an intensity proportional
to $\gamma$.  Historically, TRD's do not have the best track record,
although most problems that have been experienced arise from trying to
make TRD's that are lightweight and/or thin, so as not to degrade the
track unnecessarily before its next measurement.  Since the muon
detectors are the last element to measure a particle, there is no
incentive to make a TRD too thin.  Relatively thick detectors with
correspondingly large signals can be built.  TRD's cannot be made
arbitrarily thick, however, because a showering muon produces
electrons, which have large $\gamma$.  The TRD's would then respond to
these electrons and (correctly) identify them as high velocity
particles.

   For a VLHC operating at luminosities of $10^{34} cm^{-2}s^{-1}$,
muon identification and momentum measurement to better than 10\% for a
10 TeV muon seems possible by simple extrapolation from known
technologies.  Non-Gaussian tails causing lower $p_T$ muons to appear
as higher $p_T$ muons is a concern, which can be addressed by an
additional momentum measurement and/or a velocity measurement using
transition radiation or $dE/dx$.

    A critical issue that has not yet been addressed is that of
triggering.  Experience has shown that triggering on muons is a more
difficult problem than offline reconstruction; how much more difficult
depends on the bandwidth limitations.  Detector design may be driven
by ease of triggering.

\clearpage

\section{Conclusions}

Different scenarios of physics beyond the SM were investigated by the
physics working groups and each time the potential of the VLHC was
clearly demonstrated.  Let us review the conclusions of the different
working groups.

{\bf New Strong Dynamics Working Group}.  If strong dynamics is involved in
electroweak symmetry breaking, the physics associated with it will
first appear at the 1 TeV scale, and the VLHC will have the
opportunity to explore it in more depth than the LHC.  For example, if
a Higgs is discovered in the 400-800 GeV range, the VLHC would be able
to differentiate between a SM Higgs and a non-SM Higgs much better
than the LHC.  New strong dynamics as well as any phenomena associated
with flavor physics would also give a rich structure in the 1-10 TeV
range.  A challenge to theorists is to identify the possibilities for
10 TeV-scale physics.

{\bf Supersymmetry Working Group}.  If SUSY is discovered at low energy, as
many suspect it will be, and is gauge-mediated, one could then expect
new gauge bosons in the 10-100 TeV range.  A VLHC would be the right
place to study these new particles as well as the heavy part of the
SUSY spectrum.

{\bf Exotics Working Group}.  It is likely that not all outstanding
questions will be answered by the LHC.  Why are there three generations?  
What is the origin of the quark mixing matrix?  Are there any connections
between quarks and leptons?  We can therefore
expect some new phenomena that might manifest themselves by contact
interactions and/or by new massive particles.  The search potential of
the VLHC for these new phenomena is truly enormous and the limits are
in general in the multi-ten TeV region.  We might never get any clues
for these before the VLHC.

{\bf Full Rapidity Physics Working Group}.  A full acceptance detector will
provide a powerfull tool and investigate physics complementary to the
central, high Pt detector.  The long straight section that is needed
to insert such a detector should be included right from the start in
the VLHC design.

{\bf Precision Measurements of Heavy Objects Working Group}.  The VLHC will
not be competitive for precision measurements in the few 100 GeV mass
scale, the competition from lower energy machines is too big in that
region.  The strong suit of the VLHC is clearly its kinematic reach.

{\bf Multiple Interaction Working Group}.  The average number of
interactions per crossing at a luminosity of $10^{34}cm^{-2} s^{-1}$
should be about 22 for 17ns bunch spacing.  The situation will be
worse than, yet comparable to the situation at the LHC.  The higher
energy of the VLHC, however, will make the underlying event problem
more difficult.  This problem should be manageable if one is searching
for relatively high energy particles and jets.  The VLHC will benefit
from low-luminosity running at start-up, both for physics and detector
calibration reasons.  Luminosity callibration would also require to
operate the VLHC at the LHC energy.

{\bf Tracking Working Group}.  The total number of detector elements needed
per tracking layer is estimated to be 4x$10^{5}$ in order to keep the
occupancy down to 1\%.  To keep the momentum resolution of 10 TeV
charged leptons in the few percent range will require a tracking
resolution which is below the LHC goal, a challenging task.
Different types of tracking detectors and their potential applications
to the VLHC were discussed: two and three dimensional pixels for inner
tracking, micro strip and micro gap gas chambers for outer tracking,
and scintillating fibers.

{\bf Calorimetry Working Group}.  The calorimetry appears to be the
most feasible part of the detector.  The calorimetry radiation doses
at $E_{cm}$=100 TeV are only 2 times higher than those at the LHC,
with the same luminosity, and of course the energy resolution improves
with energy.  There are many calorimetry techniques which might fit
the VLHC requirements of good time resolution, high radiation hardness
and fine segmentation.  The in situ calibration at high energies will
be a challenge.

{\bf Muon Working Group}.  Muons are the signatures of many processes
generated by physics beyond the SM and are simple to identify.
Momentum measurement of 10 TeV muons to better than 10\% seems
possible with reasonable extrapolation from current technology.
Non-Gaussian tails causing lower $p_T$ muons to appear as higher $p_T$
muons is a concern.  This could be addressed by a second momentum
measurement using transition radiation or the (relatively large at
these energies) energy loss in a calorimeter.  Triggering will be a
serious challenge and will be limited by bandwidth consideration.
Detector design may be driven by ease of triggering.

We note that the conclusions reached during the Snowmass
96~\cite{SNOW96} workshop were confirmed by the specific studies done
during the workshop.

The VLHC will be designed to investigate the unknown physics beyond the
SM.  It should be capable of investigating a broad spectrum of models
which go beyond the SM.  Much work remains to be done even though
progress has been made during this workshop.  We can however draw the
following general conclusions.  With the center-of-mass energy and
luminosity considered at this workshop, the VLHC will be able to probe
in detail the physics that will hopefully be discovered by the LHC at
the 1 TeV mass scale.  It will furthermore allow us to investigate
scales that are about an order of magnitude larger (in some cases even larger)
than the scales probed at the LHC.  It will, however, be difficult for
the VLHC to achieve competitive measurements at the 100 GeV mass scale.
 
For a luminosity comparable to the LHC luminosity, VLHC detectors seem
feasible.  There are however many challenges and new and/or old
technologies should be pushed with the idea of decreasing the cost.
Considering the cost of LHC-like detectors, it is clear that detectors
should not be ignored in the overall cost optimization of the project.
An increase of the accelerator energy increases its cost, but allow a
decrease in luminosity (for fixed physics goal(s)) which more than
likely will decrease the cost of the detector.  The VLHC detector R\&D
effort should logically start once the LHC effort slows down.

We believe that the physics potential of a Very Large Hadron Collider
warrants a strong R\&D effort on accelerator technologies that would
enable us to reach the necessary energy and luminosity within a
reasonable cost.

Since the workshop we have started a VLHC Study Group; for more
information see~\cite{WEBPAGE}.  This workshop was sponsored by Fermi
National Accelerator Laboratory and the US Department of Energy\\

\end{document}